\documentclass[twocolumn]{aastex63}

\usepackage{natbib}
\usepackage{multirow}
\usepackage{color}
\usepackage{bm}
\usepackage[normalem]{ulem}

\newcommand{\eg}{e.g., }
\newcommand{\ie}{i.e., }
\newcommand{\Msun}{M_{\odot}}

\newcommand{\Ye}{Y_{\rm e}}
\def\gsim{\mathrel{\rlap{\lower 4pt \hbox{\hskip 1pt $\sim$}}\raise 1pt \hbox {$>$}}}
\def\lsim{\mathrel{\rlap{\lower 4pt \hbox{\hskip 1pt $\sim$}}\raise 1pt \hbox {$<$}}}

\shorttitle{Signatures of $r$-process elements in kilonova spectra}
\shortauthors{N. Domoto et al.}

\begin{document}

\title{Signatures of $r$-process elements in kilonova spectra}

\correspondingauthor{Nanae Domoto}
\email{n.domoto@astr.tohoku.ac.jp}

\author{Nanae Domoto}
\affiliation{Astronomical Institute, Tohoku University, Aoba, Sendai 980-8578, Japan}

\author{Masaomi Tanaka}
\affiliation{Astronomical Institute, Tohoku University, Aoba, Sendai 980-8578, Japan}

\author{Shinya Wanajo}
\affiliation{Max-Planck-Institut f\"{u}r Gravitationsphysik (Albert-Einstein-Institut), Am M\"{u}hlenberg 1, D-14476 Potsdam-Golm, Germany}
\affiliation{Interdisciplinary Theoretical and Mathematical Sciences Program (iTHEMS), RIKEN, Wako, Saitama 351-0198, Japan}

\author{Kyohei Kawaguchi}
\affil{Institute for Cosmic Ray Research, The University of Tokyo, 5-1-5 Kashiwanoha, Kashiwa, Chiba 277-8582, Japan}



\begin{abstract}
  Binary neutron star (NS) mergers have been expected to synthesize $r$-process elements and emit radioactively powered radiation, called kilonova.
  Although $r$-process nucleosynthesis was confirmed by the observations of GW170817/AT2017gfo, no trace of individual elements has been identified except for strontium.
  In this paper, we perform systematic calculations of line strength for bound-bound transitions and radiative transfer simulations in NS merger ejecta toward element identification in kilonova spectra. 
  We find that Sr II triplet lines appear in the spectrum of a lanthanide-poor model, which is consistent with the absorption feature observed in GW170817/AT2017gfo. 
  The synthetic spectrum also shows the strong Ca II triplet lines. 
  This is natural because Ca and Sr are co-produced in the material with relatively high electron fraction and their ions have similar atomic structures with only one $s$-electron in the outermost shell.
 The line strength, however, highly depends on the abundance distribution and temperature in the ejecta.
  For our lanthanide-rich model, the spectra show the features of doubly ionized heavy elements, such as Ce, Tb and Th.
  Our results suggest that the line forming region of GW170817/AT2017gfo was lanthanide-poor.
  We show that the Sr II and Ca II lines can be used as a probe of physical conditions in NS merger ejecta.
  Absence of the Ca II line features in GW170817/AT2017gfo implies that the Ca/Sr ratio is $< 0.002$ in mass fraction, which is consistent with nucleosynthesis for electron fraction $\ge 0.40$ and entropy per nucleon (in units of Boltzmann constant) $\ge 25$. 
\end{abstract}



\section{Introduction}
\label{sec:intro}
Coalescence of binary neutron stars (NSs) is a promising site for the rapid neutron capture nucleosynthesis ($r$-process, \eg \citealp{LattimerSchramm1974, Eichler1989, Freiburghaus1999, Goriely2011a, Wanajo2014}). 
A NS merger ejects neutron-rich material and heavy elements are synthesized in the ejecta. 
Then, radioactive decay of freshly synthesized nuclei powers electromagnetic emission, so called kilonova \citep{LiPaczynski1998, Metzger2010, Roberts2011, Korobkin2012}. 
A kilonova is expected to produce thermal emission mainly in ultraviolet (UV), optical and near infrared (NIR) wavelengths \citep{Kasen2013, BarnesKasen2013, TanakaHotokezaka2013}. \par

NS mergers can generate several ejecta components reflecting the variety of mass ejection mechanisms. 
One is the dynamical ejecta component, which is promptly ejected after the merger with high velocity ($v \sim 0.2\ c$, where $c$ is the speed of light) by tidal torque and shock heating (\eg \citealp{Rosswog1999, Hotokezaka2013, Sekiguchi2015}). 
Another is the post-merger ejecta component from a subsequently formed accretion disk, which is driven by viscosity heating (\eg \citealp{FernandezMetzger2013, Just2015, Fujibayashi2018, Fujibayashi2020}). 
Mass ejection can be also enhanced by neutrino heating if the massive NS remnant survives (\eg \citealp{Ruffert1997, Perego2014, Fujibayashi2020a}). \par

NS mergers are also the primary targets of gravitational wave (GW) observations.
In fact, the first GW detection from a NS merger was successfully made in 2017 (GW170817, \citealp{Abbott2017a}). 
Through the intensive follow-up observations, an associated electromagnetic counterpart, AT2017gfo, was also identified (\citealp{Abbott2017b}). 
Observed properties of AT2017gfo in the UV, optical and NIR wavelengths are broadly consistent with the theoretical expectation of kilonova (\eg \citealp{Arcavi2017, Coulter2017, Evans2017, Pian2017, Smartt2017, Utsumi2017, Valenti2017}). 

It is important to figure out the abundance patterns which NS mergers produce. 
This leads toward understanding of not only the origin of heavy elements but also the physical conditions for NS merger ejecta, \ie their masses, velocities and electron fractions (the number of protons per nucleon, $\Ye$). 
When the ejecta have low $\Ye$ ($\lesssim 0.25$), a strong $r$-process takes place and heavy elements including lanthanides ($Z=57$--71) can be synthesized. 
Since lanthanides have high opacity in the ejecta, a kilonova is expected to show a long-lasting emission in the NIR wavelength (``red" component, \citealp{Kasen2013, TanakaHotokezaka2013}). 
On the other hand, if $\Ye$ in the ejecta is relatively high ($\Ye \gtrsim 0.25$), the production of heavy elements is suppressed and lighter $r$-process elements can be synthesized \citep{MetzgerFernandez2014}. 
Such ejecta can give rise to a short-time kilonova which is bright in the optical wavelength (``blue" component, \citealp{Kasen2015, Tanaka2018}). 
The time evolution of luminosity and color of AT2017gfo suggests the presence of both ``red" and ``blue" components (\eg \citealp{Kasen2017, Perego2017, Shibata2017, Tanaka2017, Kawaguchi2018, Rosswog2018}). 
This has provided us with the evidence that NS mergers can be certainly the site of $r$-process nucleosynthesis. 

It is not yet clear, however, which elements are synthesized in this NS merger. 
It is challenging to identify individual elements by using absorption features in the observed spectra since the absorption lines become broader and significantly overlapped in the high velocity ejecta. 
In fact, no elements have been identified except for Sr in the spectra of GW170817/AT2017gfo \citep{Watson2019}. 
To extract elemental information from kilonova spectra, 
we need to understand which elements produce significant absorption lines in kilonova spectra by means of detailed theoretical calculations.

In this paper, we perform systematic calculations of line strength for bound-bound transitions and radiative transfer simulations in NS merger ejecta toward element identification in kilonova spectra. 
In Section \ref{sec:line}, we show the line strength of elements for a simple one zone model. 
Then, we perform radiative transfer simulations by assuming a one-dimensional ejecta structure in Section \ref{sec:spectra}. 
We also discuss possible absorption features caused by other heavy elements in Section \ref{sec:discussion}. 
Finally we give our conclusions in Section \ref{sec:conclusion}.

\section{Systematic calculations of line strength}
\label{sec:line}
\subsection{Methods}
\label{sec:line method}
\begin{figure*}[ht]
  \begin{center}
    \begin{tabular}{cc}
    \includegraphics[scale=0.7]{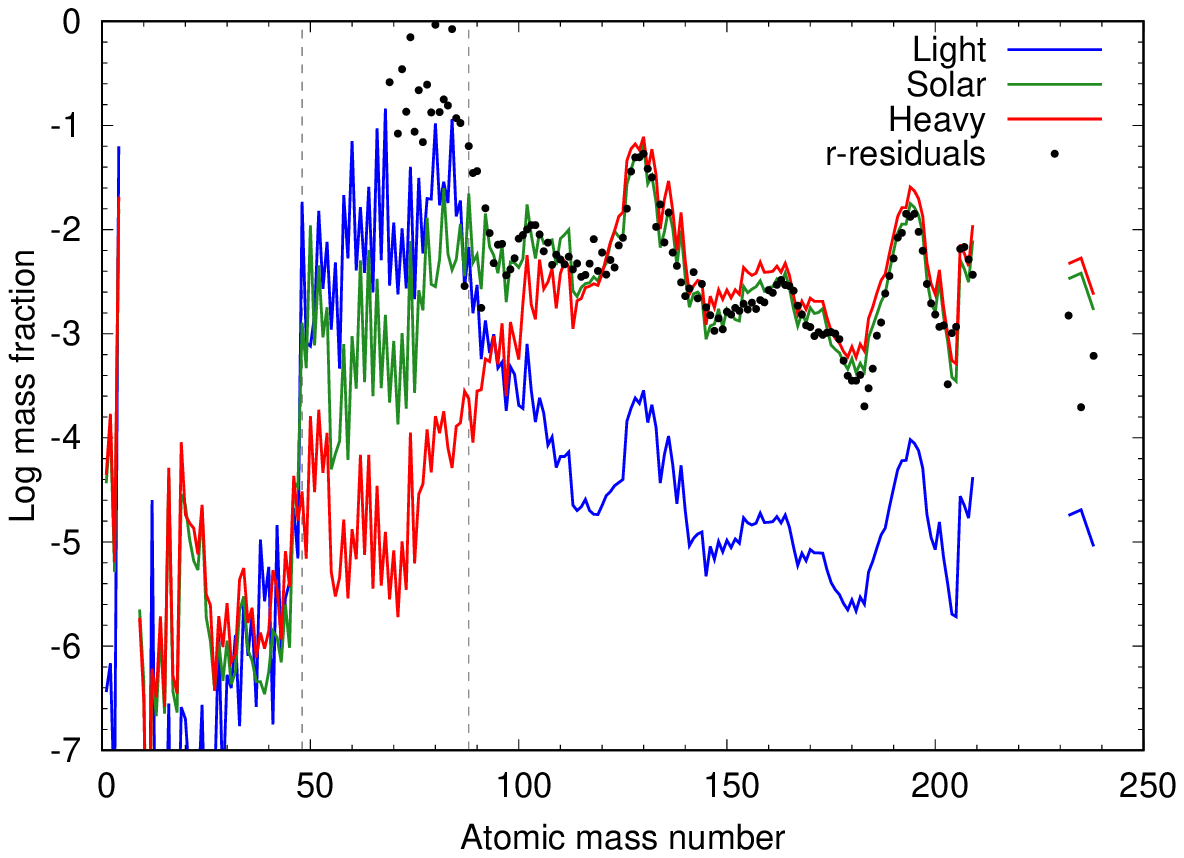} &
    \includegraphics[scale=0.7]{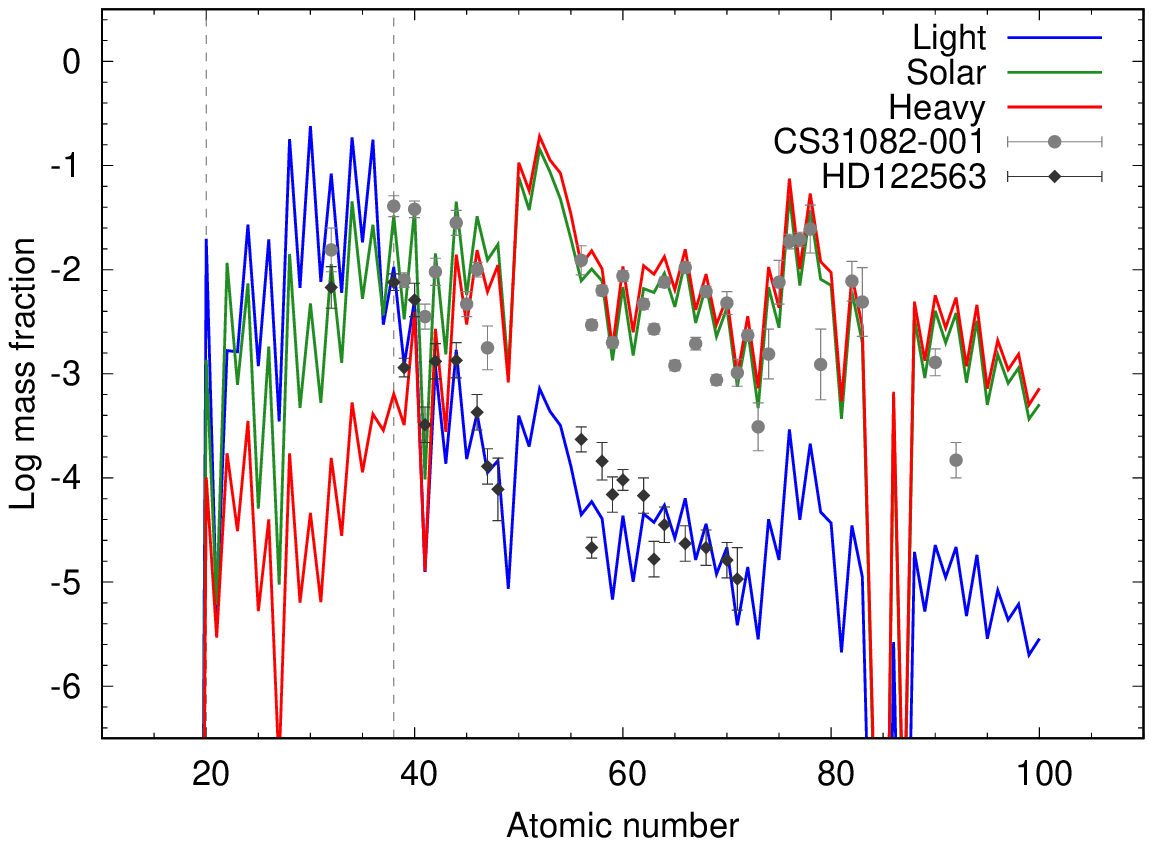}  \\
    \end{tabular}
\caption{
  \label{fig:abun}
  Left: Final abundances for each model as a function of mass number. 
  Black dots show the $r$-process residual pattern used for fitting \citep{Prantzos2020}. 
  The residual abundances are scaled to match those for the S model at $A=138$. 
  Vertical dashed lines indicate $^{48}$Ca and $^{88}$Sr. 
  Right: Abundances for each model at $t=1.5$ days as a function of atomic number. 
  Abundances of an $r$-process-enhanced star CS 31082-001 (circles, \citealp{Mello2013}) and an $r$-process-deficient star HD 122563 (diamonds, \citealp{Honda2006}; Ge from \citealp{Cowan2005}; Cd and Lu from \citealp{Roederer2012}) are also shown for comparison purposes. 
  The abundances of CS 31082-001 and HD 122563 are scaled to those for the S and L models at $Z=40$, respectively. 
  Vertical dashed lines indicate Ca ($Z=20$) and Sr ($Z=38$). 
}
\end{center}
\end{figure*}
\begin{figure}
  \begin{center}
    \includegraphics[scale=0.68]{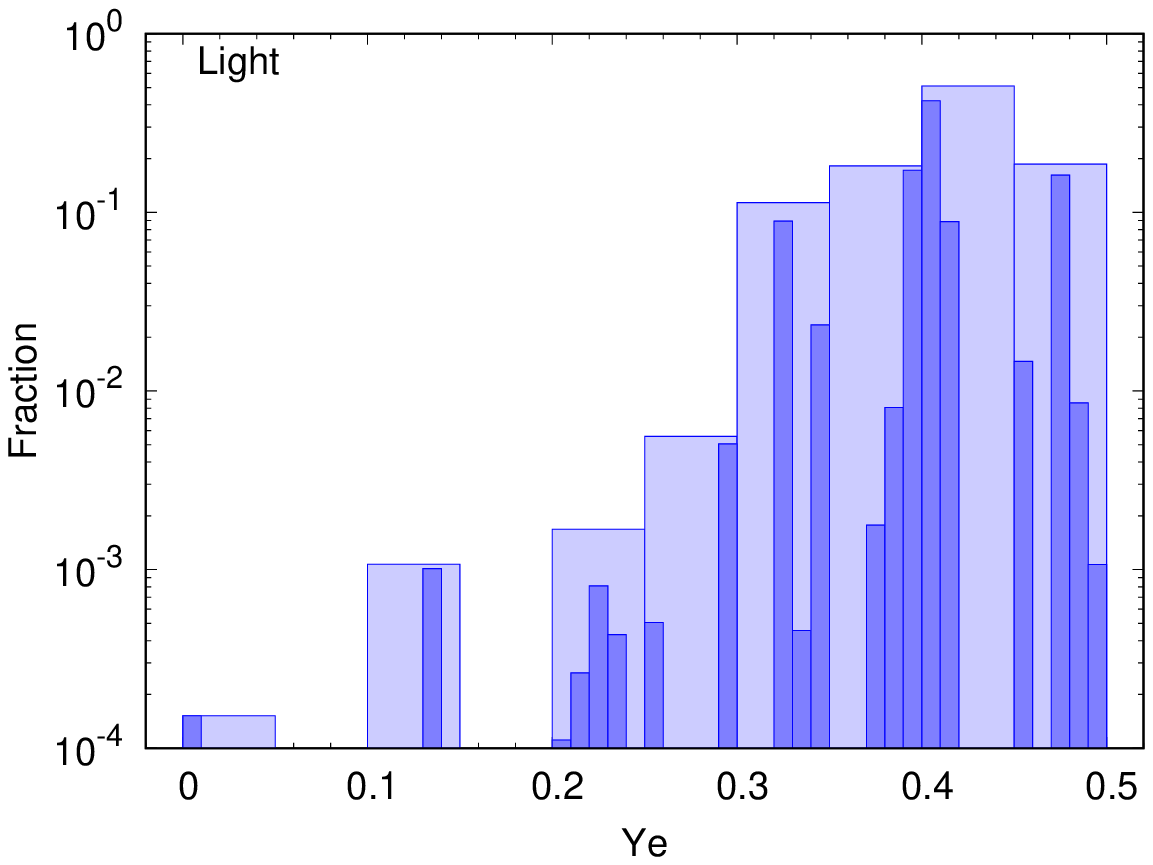} \\
    \includegraphics[scale=0.68]{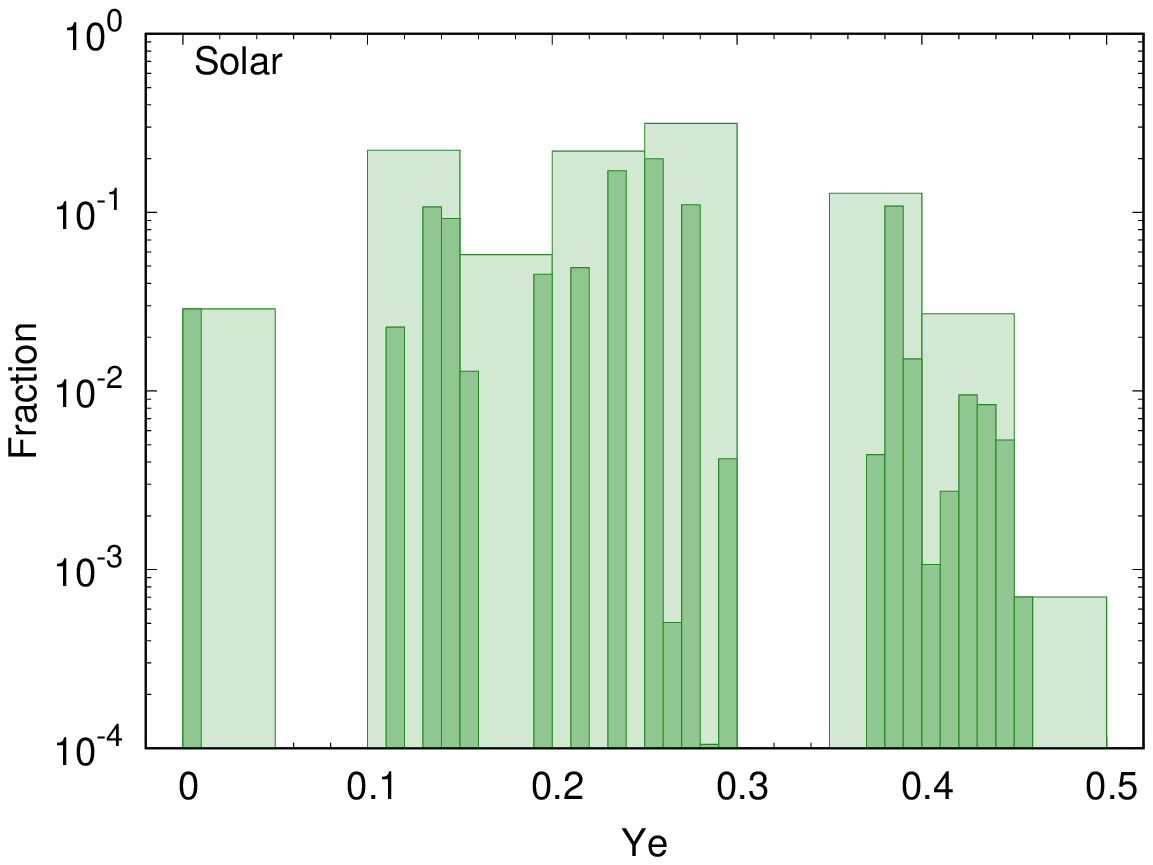} \\
    \includegraphics[scale=0.68]{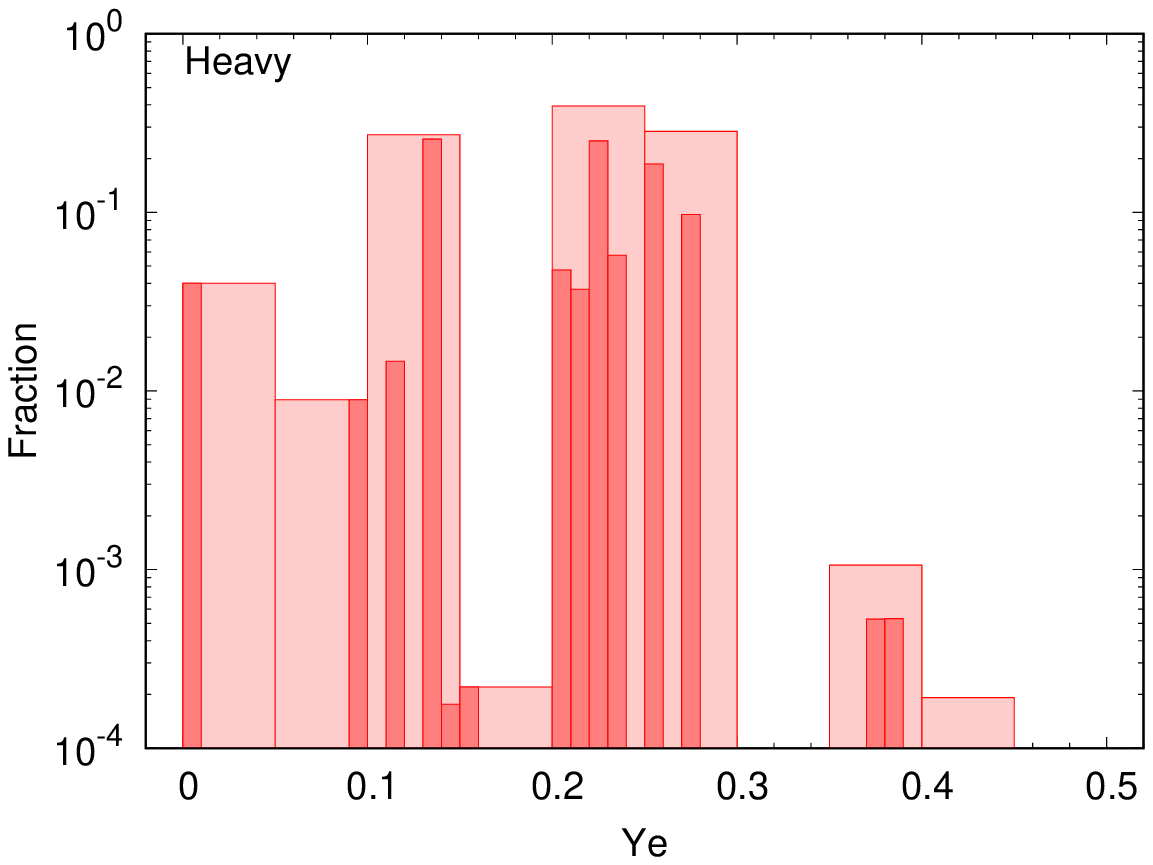}  \\
\caption{
  \label{fig:hist ye}
  Histograms of electron fraction $\Ye$ 
  for the L (top), S (middle) and H (bottom) models. 
  Denser colors show the histograms with an original interval ($\Delta\Ye=0.01$), while lighter colors show those with a grouped interval ($\Delta\Ye=0.05$).
}
\end{center}
\end{figure}
To investigate which elements produce strong absorption lines in the kilonova spectra, 
we systematically calculate the strength of bound-bound transitions for given density, temperature and element abundances. 
The line strength is approximated by the Sobolev optical depth \citep{Sobolev1957} for each bound-bound transition, 
\begin{equation}
	\tau_l = \frac{\pi e^2}{m_e c} f_l n_{i, j} t \lambda_l
	\label{eq:tau}
\end{equation}
for homologously expanding ejecta. 
The Sobolev approximation is valid for expanding matter with a large radial velocity gradient. 
Here $n_{i, j}$ is the number density of the lower level of the transition ($i$-th ionized element in the $j$-th excited state), $f_l$ and $\lambda_l$ are the oscillator strength and the transition wavelength of the bound-bound transition, respectively. 
As in previous work on kilonova (\eg \citealp{BarnesKasen2013, TanakaHotokezaka2013}), we assume local thermodynamic equilibrium (LTE); we assume Boltzmann distribution for the population of excited levels and solve the Saha equation to obtain ionization states.  \par

Atomic data are essentially important to evaluate the line strength.
For theoretical calculations of kilonova light curves, atomic data from theoretical calculations have been often used (\eg \citealp{Kasen2013, Tanaka2018, Fontes2020, Banerjee2020}). 
This is useful in terms of completeness of data because the opacity of the ejecta should be correctly evaluated for the light curve calculations (see Appendix \ref{sec:appendix1}). 
However, while such theoretical data may give a reasonable estimate for the total opacity, they are not accurate in transition wavelengths, and thus, not suitable for element identification. 
In this work, since we focus on the imprints of elemental abundances in kilonova spectra, 
we construct the latest line list based on the Vienna Atomic Line Database (VALD; \citealp{Piskunov1995, Kupka1999, Ryabchikova2015}). 
This database is suitable to identify lines because the transition wavelengths are calibrated with experiments and semi-empirical calculations. 
It should be, however, noted that the line list is not necessarily complete in particular in the NIR region, although kilonovae are brighter in the NIR at late times. 
The impact of the incompleteness is discussed in Section \ref{sec:NIR}. \par 

For the abundance in the ejected matter from a NS merger, we use a multi-component free-expansion (mFE) model in \citet{Wanajo2018a}. 
This model is constructed as an ensemble of the parameterized outflows with constant velocity, initial entropy, and initial $\Ye$, which fit the $r$-process residuals of the solar abundances \citep{Prantzos2020}. 
The ranges of velocity (in units of $c$), entropy (in units of Boltzmann constant per nucleon, $k_{\rm B}$/nuc), and $\Ye$ are taken to be 0.05--0.30, 10--35, and 0.01--0.50 with the intervals of 0.05, 5, and 0.01, respectively, as in \citet{Wanajo2018a}. 
Here we consider three models (the left panel of Figure \ref{fig:abun}): 
(1) a model that fits the $r$-process residuals for $A \ge 88$, where $A$ is mass number, \ie including those heavier than the first $r$-process peak isotopes ($A = 80$--84),
(2) a model that fits those for $A \ge 69$ (including the first $r$-process peak isotopes) and 3\% of those for $A\ge 100$, and 
(3) a model that fits those for $A\ge 88$ and 1\% of those for $A < 110$. 
Hereafter we refer to each model as the Solar (S), Light (L), and Heavy (H) model, respectively. 
Note that the S model is the same as mFE-b in \citet{Wanajo2018a} but the $r$-process residuals have been updated to those in \citet{Prantzos2020}. 
The L model exhibits a similar abundance pattern to that of a metal-poor star with weak $r$-process signature (\eg HD122563, \citealp{Honda2006}; the right panel of Figure \ref{fig:abun}). 
The H model represents a putative case, \eg with a contribution from only dynamical ejecta. 
Note that the minimum mass number $A=88$ for the S and H models corresponds to the dominant isotope of Sr, the element that has been measured in all $r$-process-enhanced stars (\eg \citealp{Cowan2019}).
We also performed the same calculations with the minimum mass number replaced by $A=85$ (excluding the first $r$-peak and lighter isotopes) and 90 and confirmed that our results are unaffected by this choice. \par 

Although these models include the abundances with $Z=1$--110, we use only those with $Z=20$--100 in our calculations at $t=1.5$ days as shown in the right panel of Figure \ref{fig:abun}. 
The heaviest elements with $Z \ge 101$ are excluded because their mass fractions are very small ($\sim 10^{-6}$--$10^{-4}$) and there is no atomic data for such heavy elements. 
The light elements with $Z\le 19$ are also excluded 
because their mass fractions are also small, an order of $10^{-4}$, and they do not affect our results (see also \citet{Perego2020} for the effects of lightest elements). 
We summarize the mass fractions of selected elements for our models in Table \ref{tab:abun}. 
The distributions of $\Ye$ for these models are shown in Figure \ref{fig:hist ye} (see also Appendix \ref{sec:appendix2} for the distributions of velocity and entropy). 
As can be anticipated from the abundance patterns (Figure \ref{fig:abun}), the distributions for the L and H models are dominated by higher ($>$ 0.3) and lower ($<$ 0.3) values of $\Ye$ than those for the S model.

\begin{deluxetable}{llll}[ht]
\tablewidth{0pt}
\tablecaption{Mass fractions of selected elements. 
 The top rows show the final abundances and the bottom rows show those at $t=1.5$ days for each model.}
\label{tab:abun}
\tablehead{
  Model   & X(Ca)$^a$ & X(Sr)$^b$ & X(La+Ac)$^c$ 
}
\startdata
Light (L)     & $1.8\times 10^{-2}$ & $6.6\times 10^{-3}$ & $4.9\times 10^{-4}$ \\ 
		& $2.0\times 10^{-2}$ & $1.1\times 10^{-2}$ & $5.9\times 10^{-4}$ \\ \hline 
Solar (S)    & $1.3\times 10^{-3}$ & $2.2\times 10^{-3}$ & $0.09$ \\ 
		& $1.4\times 10^{-3}$ & $3.3\times 10^{-2}$ & 0.10 \\ \hline
Heavy (H)  & $8.6\times 10^{-5}$ & $2.4\times 10^{-4}$ & $0.13$ \\
		& $1.0\times 10^{-4}$ & $6.4\times 10^{-4}$ & 0.15 \\
\enddata
\tablecomments{
$^a$ Mass fraction of calcium ($Z=20$). \\
$^b$ Mass fraction of strontium ($Z=38$). \\
$^c$ Sum of mass fractions for lanthanides ($Z=57$--71) and actinides ($Z=89$--100).
}
\end{deluxetable}

\subsection{Results}
\label{sec:line result}
\begin{figure*}[th]
  \begin{center}
    \begin{tabular}{cc}
    \includegraphics[scale=0.7]{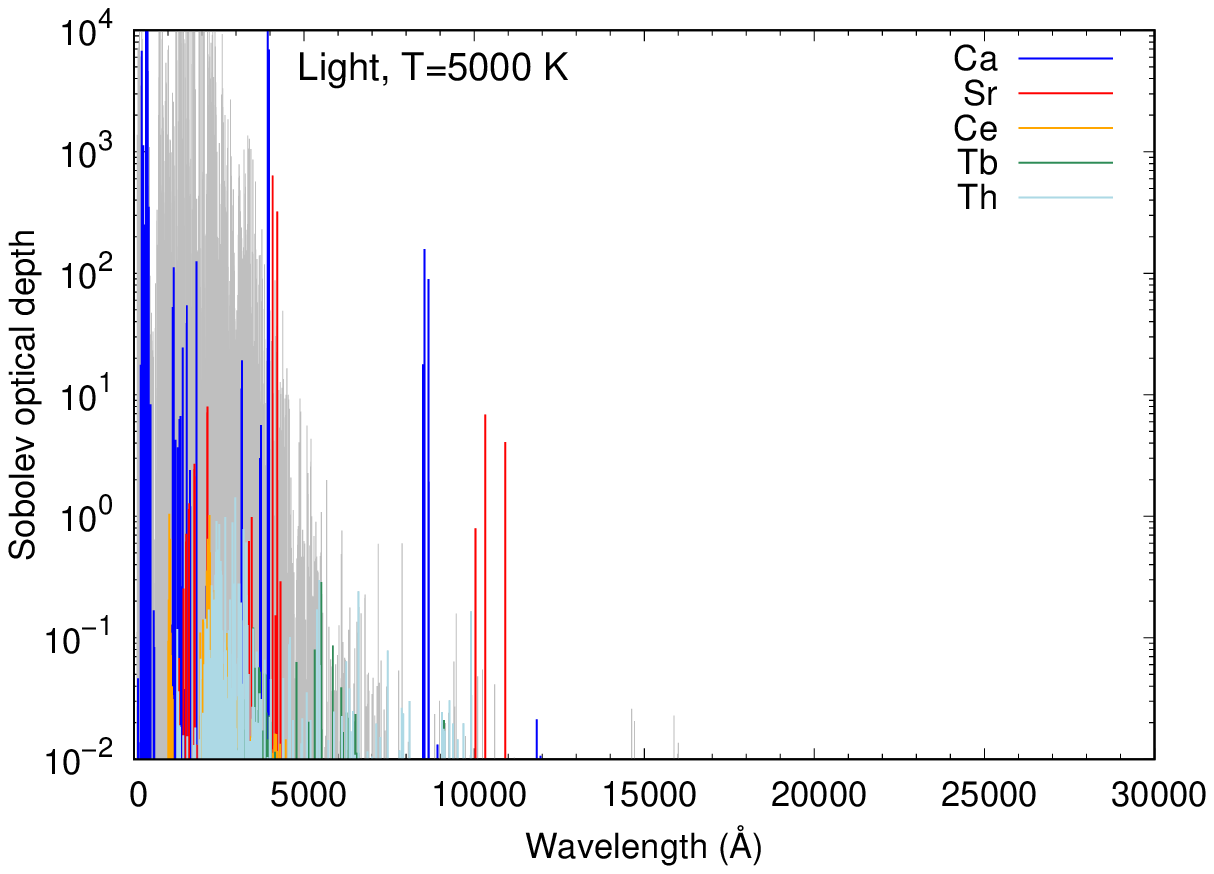} &
    \includegraphics[scale=0.7]{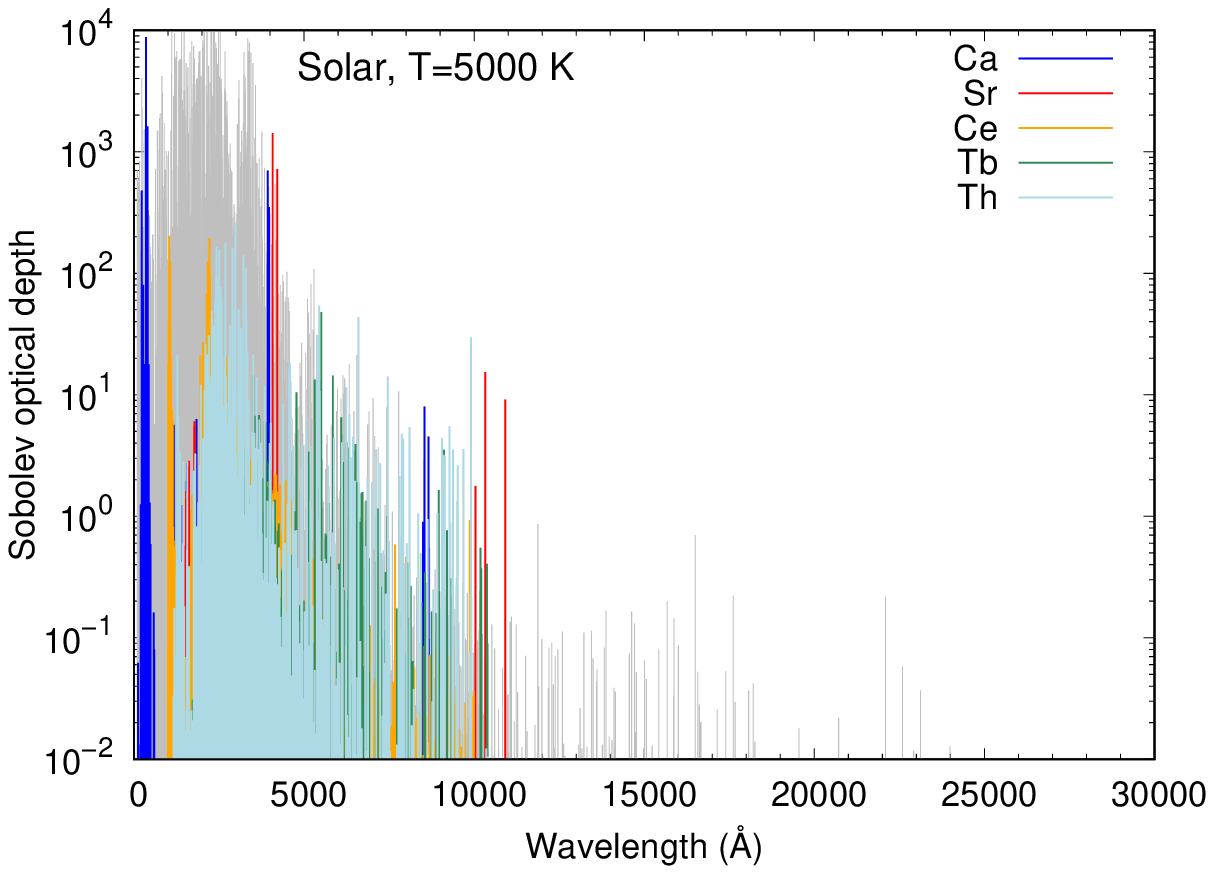}  \\
    \includegraphics[scale=0.7]{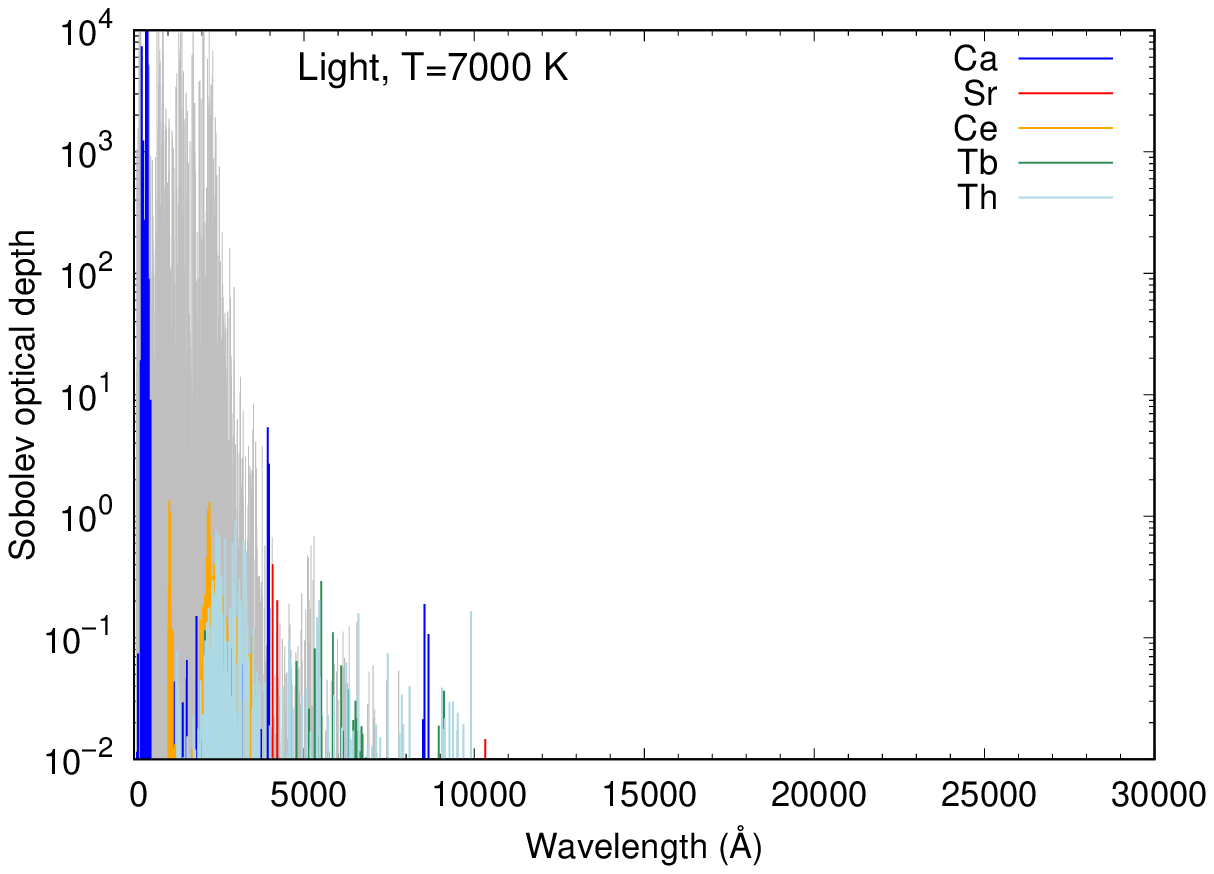} &
    \includegraphics[scale=0.7]{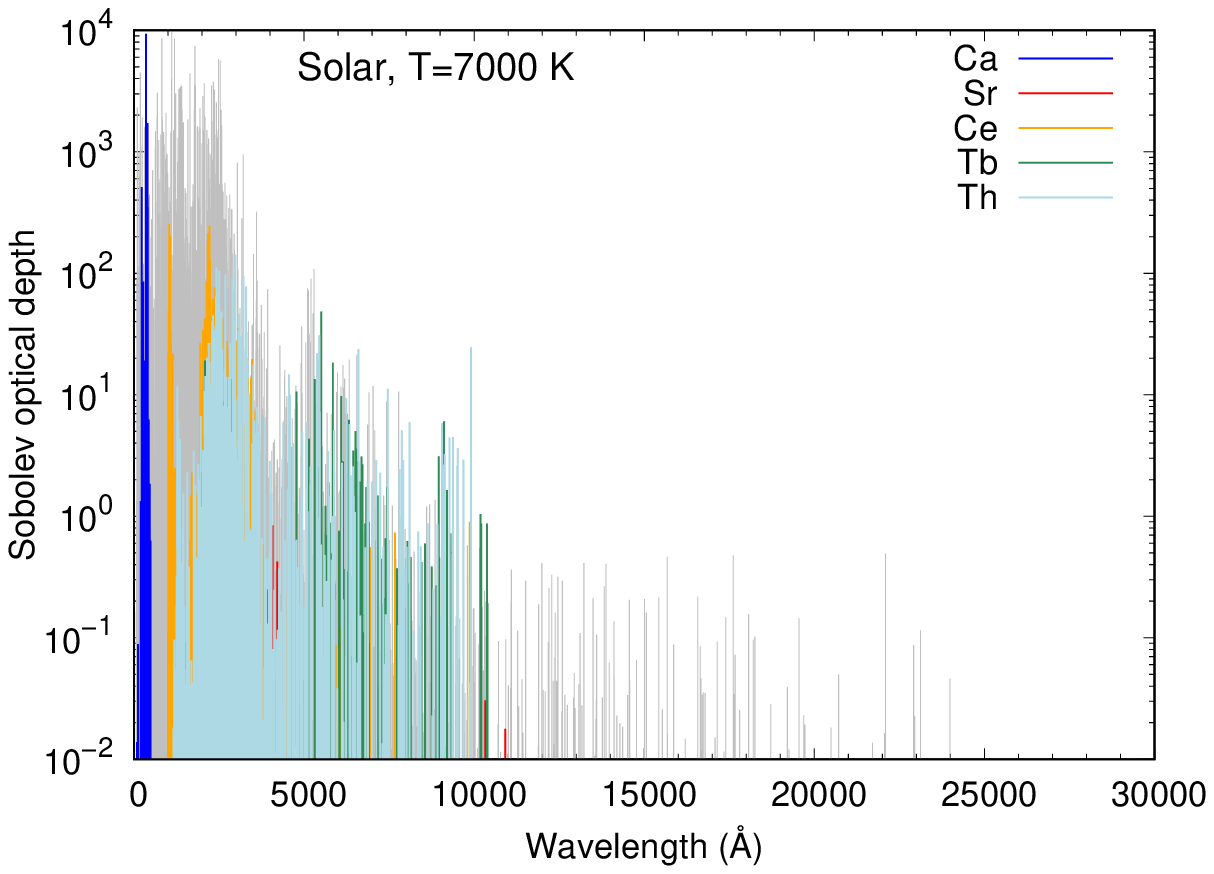}  \\
    \end{tabular}
\caption{
  \label{fig:tau}
  Sobolev optical depth of bound-bound transitions at $t=1.5$ days. 
  The elements with large contributions are shown with colors. 
  The left panels show the results for the L model, while the right panels show those for the S model. 
  The top and bottom panels correspond to two different temperatures, $T=5000\ {\rm K}$ and $T=7000\ {\rm K}$, respectively. 
  For all the cases, the Sobolev optical depths are calculated with the density of $\rho=10^{-14}\ {\rm g\ cm^{-3}}$.
}
\end{center}
\end{figure*}
\begin{figure*}[ht]
  \begin{center}
    \begin{tabular}{cc}
    \includegraphics[scale=0.7]{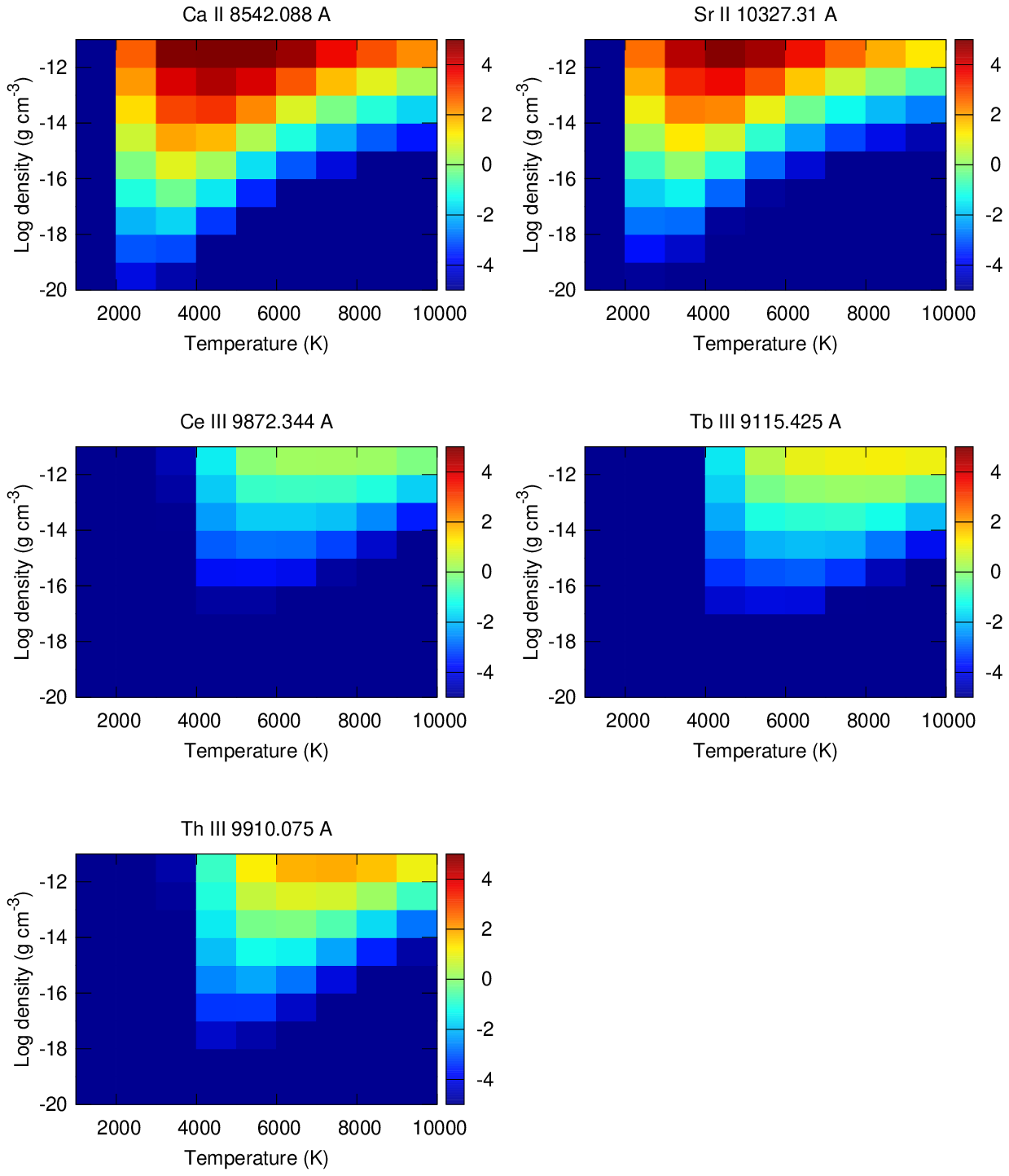} &
    \includegraphics[scale=0.7]{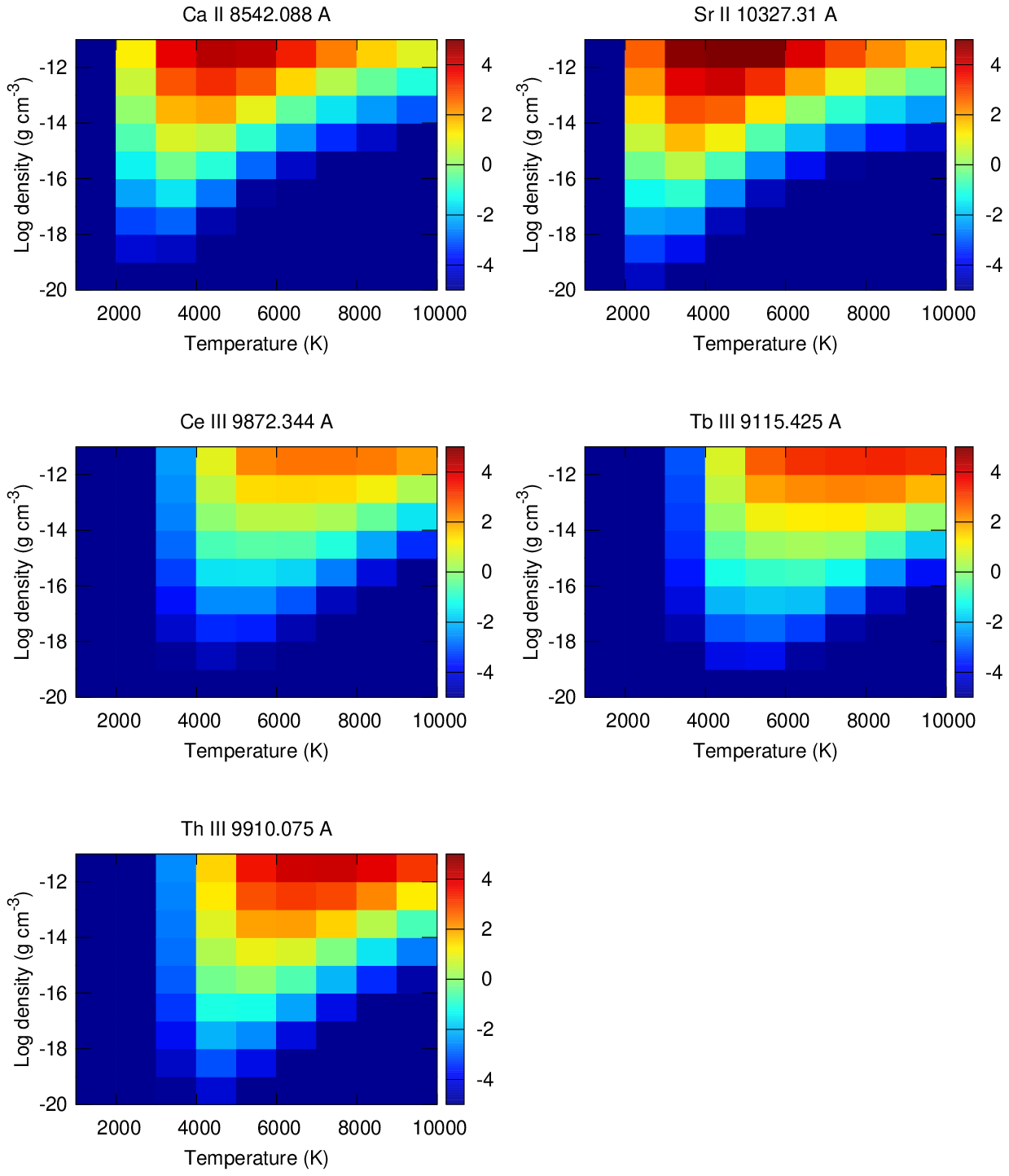}  \\
    \end{tabular}
\caption{
  \label{fig:map}
  Density and temperature dependences of line strength at $t = 1.5$ days for representative transitions.
  Colors represent the Sobolev optical depth in logarithmic scale. 
  The left five panels show the results for the abundance of the L model, while the right five panels show those of the S model. 
  For both of the Ca II and Sr II triplet lines, we show the strength of the middle transitions. 
}
\end{center}
\end{figure*}

The strength of bound-bound transitions at $t=1.5$ days is displayed in Figure \ref{fig:tau}.
As typical values in the ejecta (see Section \ref{sec:spectra}), we evaluate the Sobolev optical depth for the density of $\rho=10^{-14}\ \rm{g\ cm^{-3}}$ and the temperature of $T=5000$ K (top panels) and 7000 K (bottom panels). \par

We find that the Sr II triplet lines (red lines at 10000--11000 \AA) are strong for the L model in the case of $T=5000\ {\rm K}$. 
Other notable features are the Ca II triplet lines (blue lines around 8500 \AA), which are stronger than the Sr II lines. 
Note that Ca and Sr abundances in this model are dominated by $^{48}$Ca and $^{88}$Sr (see the left panel of Figure \ref{fig:abun}). 
For higher temperature ($T=7000\ {\rm K}$, bottom panel), these lines significantly become weaker. \par

In the S model, which has more heavy elements, either of the Sr II (red) or Ca II (blue) lines are not necessarily as strong as other lines of heavy elements such as Tb III and Th III when assuming $T=5000\ {\rm K}$. 
For higher temperature ($T=7000\ {\rm K}$), the Sr II and Ca II lines become significantly weaker as in the L model, 
while the strength of the Tb III and Th III lines is not largely affected. 
The results for the H model are similar to those for the S model. \par

The behavior of the line strength can be understood as the dependence of the Sobolev optical depths on temperature and density as well as abundance distribution. 
Figure \ref{fig:map} shows the Sobolev optical depth for ranges of the density $\rho=10^{-20}$--$10^{-11}\ {\rm g\ cm^{-3}}$ and temperature $T=1000$--10000 K. 
For the Ca II and Sr II lines, the strength of the middle line of each triplet (8452.088 {\AA} for Ca II and 10327.31 {\AA} for Sr II) is shown. 
Regarding Ce III lines, we choose one of the lines (9872.344 {\AA}) as representative. 

When we focus on the case with the density of $\rho = 10^{-14}\ {\rm g\ cm^{-3}}$ as in Figure \ref{fig:tau}, 
the strength of Ca II and Sr II line is maximal around $T \sim 4000$ K.
Then, the lines become weaker for higher temperature as these elements are ionized to the doubly ionized state. 
This is the reason why the models with $T=5000$ K show the stronger Sr II and Ca II lines compared to those with $T=7000$ K.
On the other hand, the strength of the Ce III, Tb III, and Th III lines is maximal around $T \sim 6000$ K, and thus, they do not show a large difference between $T=5000$ K and $T=7000$ K.
As a result, the line strength of these heavy elements becomes much more dominant than that of the Sr and Ca lines for $T= 7000$ K for the S and H models, in which these heavy elements are abundant.
\par

\begin{figure}[ht]
  \begin{center}
    \includegraphics[scale=0.45]{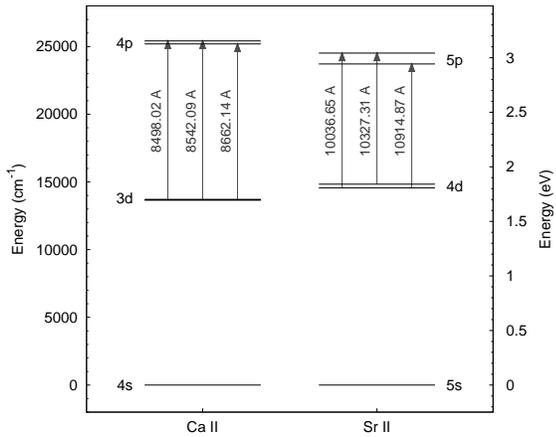}
\caption{
  \label{fig:level}
  Energy diagrams for Ca II and Sr II. 
  Each arrow shows the triplet transition with the value of the transition wavelength. 
  The energy terms for these triplet lines are 
  {\rm ${}^2D_{3/2}$--${}^2P^o_{3/2}$},
  {\rm ${}^2D_{5/2}$--${}^2P^o_{3/2}$}, and 
  {\rm ${}^2D_{3/2}$--${}^2P^o_{1/2}$} 
  from the shorter to longer wavelengths.
  }
\end{center}
\end{figure}
\begin{figure}[ht]
  \begin{center}
    \includegraphics[scale=0.66]{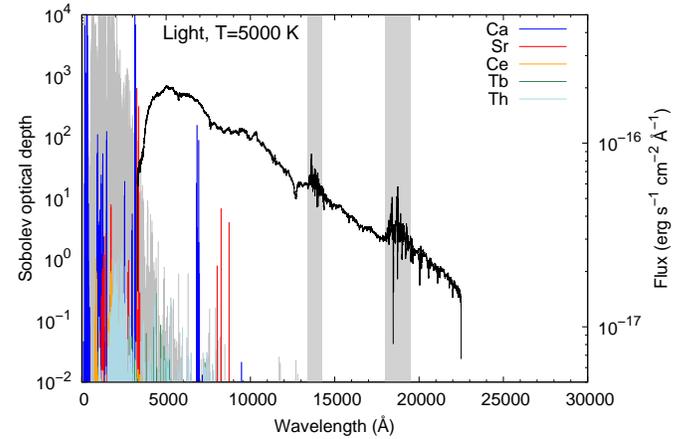}
\caption{
  \label{fig:gw-tau}
  Comparison between the observed spectrum for AT2017gfo at $t = 1.5$ days \citep{Pian2017} 
  and the calculated Sobolev optical depth for the L model ($\rho=10^{-14}\ {\rm g\ cm^{-3}}$ and $T=5000\ {\rm K}$). 
  The positions of lines are blueshifted owing to the adopted expansion velocity of $v = 0.2\ c$. 
  The gray shade shows the region of strong atmospheric absorption.
  }
\end{center}
\end{figure}

It is natural that not only Sr II but also Ca II triplet lines become strong, because these elements have similar atomic structures.
Figure \ref{fig:level} shows the energy levels and the transition wavelengths for these ions (NIST ASD, \citealp{NIST_ASD}). 
Both elements belong to the group 2 in the periodic table and have only one electron in the outermost shell when they are singly ionized. 
The electron occupies the $s$ shell in the ground state and has a relatively small number of excited levels. 
Therefore, the transition probability of each bound-bound transition tends to be high. 

We confirm that the observed absorption feature around $\lambda \sim$ 8000 {\AA} in AT2017gfo is consistent with the Sr II triplet. 
Figure \ref{fig:gw-tau} shows comparison between the spectrum of AT2017gfo at $t=1.5$ days \citep{Pian2017} 
and the result of the L model ($\rho =10^{-14}\ {\rm g\ cm^{-3}}$, $T=5000\ {\rm K}$).
In this model, the wavelengths of the transitions are blueshifted owing to the adopted expansion velocity of $v=0.2\ c$ which is suggested by the observations of AT2017gfo (\eg \citealp{Pian2017, Smartt2017}). 
With this velocity, the wavelengths of the Sr II triplet (red) lines match the observed broad absorption feature, as also demonstrated by \citet{Watson2019}. 
Note that \citet{Watson2019} included elements with $Z\ge 33$ in their calculations, \ie Ca ($Z=20$) was not included. 
Although our L model suggests that Ca II can also exhibits strong absorption lines, the spectrum of AT2017gfo does not show such a feature at $\lambda \sim$ 6800 {\AA}, an expected wavelength of the Ca II triplet with $v=0.2\ c$. 
Implications of the absence of the Ca II lines in the observed spectrum are discussed in Section \ref{sec:Ca}. \par

\section{Synthetic spectra}
\label{sec:spectra}
\subsection{Methods}

In this section, we calculate realistic synthetic spectra of kilonovae.
The analysis in the previous section has evaluated the line strength for one-zone models.
However, the ejecta from NS mergers have spatial distribution in density and temperature. 
Also, the temperature structure is controlled by radioactive heating and radiative transfer. 
Therefore, we perform radiative transfer simulations by taking the ejecta structure and radioactive heating into account. 
We use a wavelength-dependent radiative transfer simulation code \citep{TanakaHotokezaka2013, Tanaka2014, Tanaka2017, Tanaka2018, Kawaguchi2018}. 
The photon transfer is calculated by the Monte Carlo method. 
To compute the opacity for bound-bound transitions, we adopt the expansion opacity \citep{Karp1977} and use the formula from \citet{EastmanPinto1993}:
\begin{equation}
	\kappa_{\rm{exp}}(\lambda)=\frac{1}{c t \rho} \sum_l \frac{\lambda_l}{\Delta \lambda}(1-\mathrm{e}^{-\tau_l}) ,
\end{equation}
where $\tau_l$ is the Sobolev optical depth (Equation (\ref{eq:tau})). In the equation, the summation is taken over all the transitions within the wavelength bin $\Delta \lambda$. 
The Sobolev optical depths are evaluated in the same manner in Section \ref{sec:line}: we use the line list from the VALD database and assume LTE for ionization and excitation. 
Hereafter, we focus on the spectra at 1.5 days after the merger because insufficient atomic data in the NIR region start to affect the emission at later phases (see Section \ref{sec:NIR} and Appendix \ref{sec:appendix1}).

For the ejecta density structure, 
we assume a single power law  ($\rho \propto r^{-3}$) for the velocity range of the ejecta $v=0.05$--$0.3\ c$.
The total ejecta mass is set to be $M_{\rm ej}=0.03\Msun$. 
We use the three models for the abundance distributions as described in Section \ref{sec:line method}.
The heating rate of radioactive nuclei is consistently taken from each model.
The thermalization efficiency of $\gamma$-rays and radioactive particles follows the analytic formula by estimating characteristic timescales \citep{Barnes2016}. \par

\subsection{Results}
\label{sec:spectra result}
\begin{figure}[th]
  \begin{center}
    \includegraphics[scale=0.67]{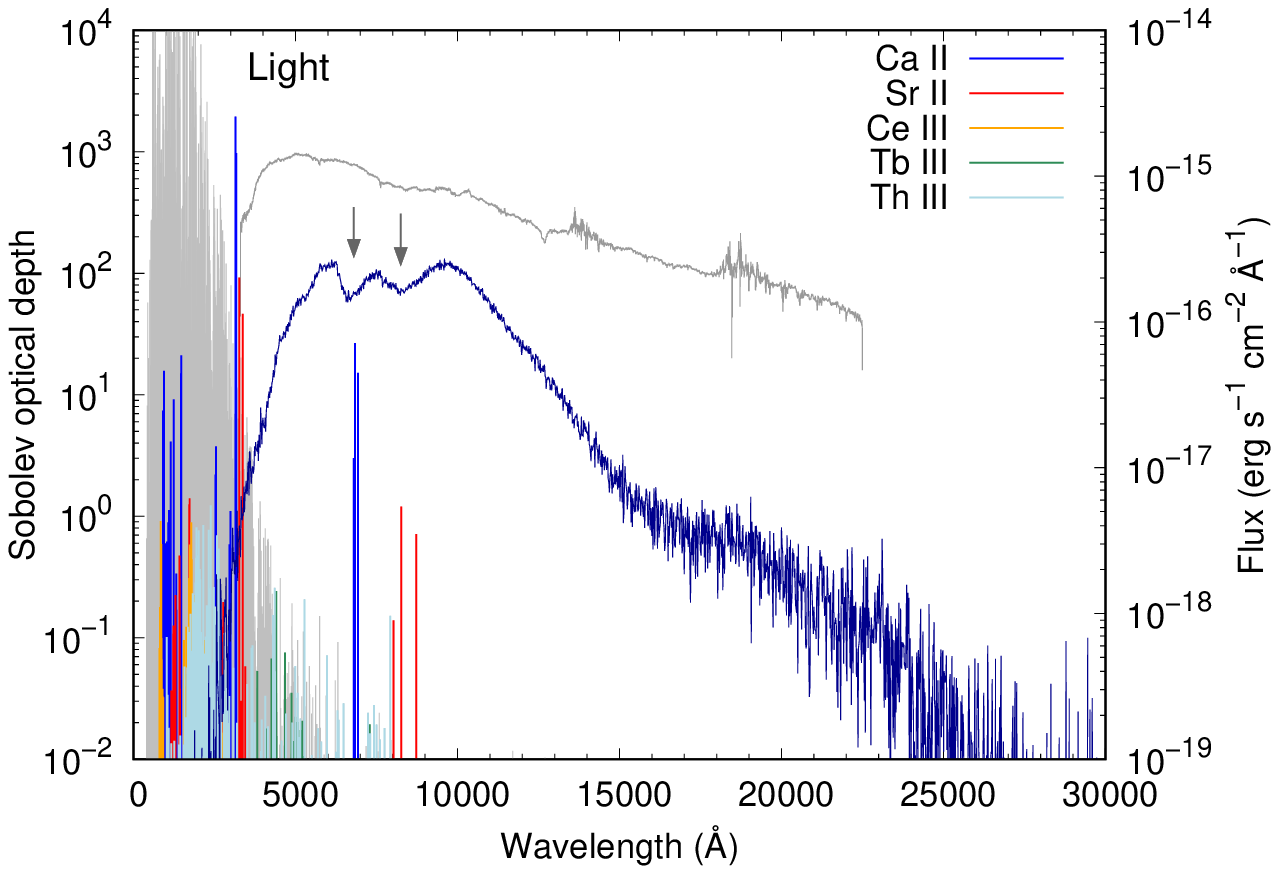} \\
    \includegraphics[scale=0.67]{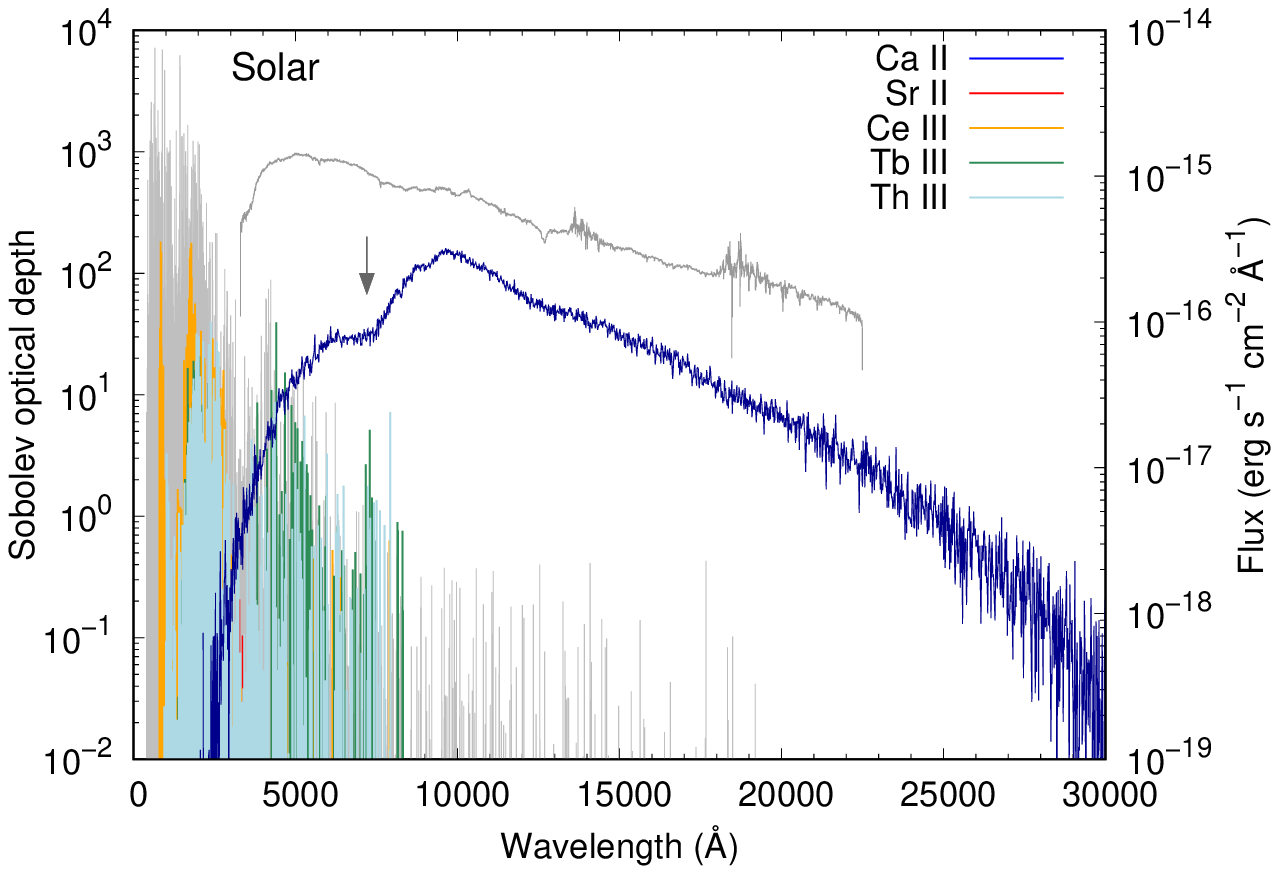}  \\
    \includegraphics[scale=0.67]{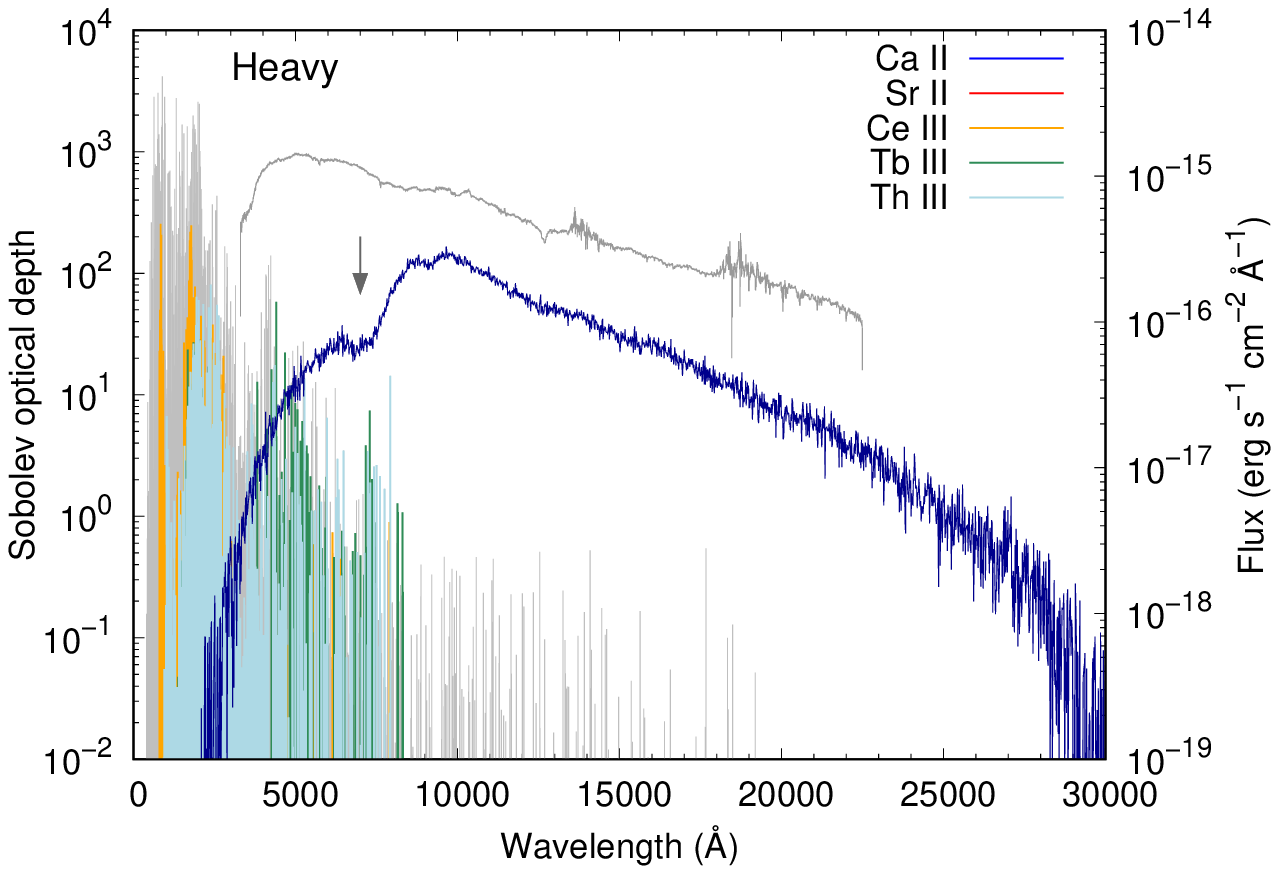}
\caption{
  \label{fig:flux}
  Synthetic spectra (blue curve) and line strength of each transition (vertical lines) at $t=1.5$ days 
  for the L (top), S (middle), and H (bottom) models.
  We plot the Sobolev optical depths in the ejecta at $v = 0.2\ c$.  
  The positions of lines are blueshifted according to $v=0.2\ c$. 
  Gray curve in each panel shows the spectrum of AT2017gfo at $t=1.5$ days, which is vertically shifted for comparison. 
  Each arrow indicates the absorption lines which we focus on (see the text). 
  The temperature in the ejecta at $v = 0.2\ c$ is $T \sim 5200\ {\rm K}$ for the L model, while it is $T\sim 7200\ {\rm K}$ for the S and H models.
}
\end{center}
\end{figure}
\begin{figure*}[thb]
  \begin{center}
    \begin{tabular}{cc}
    \includegraphics[scale=0.7]{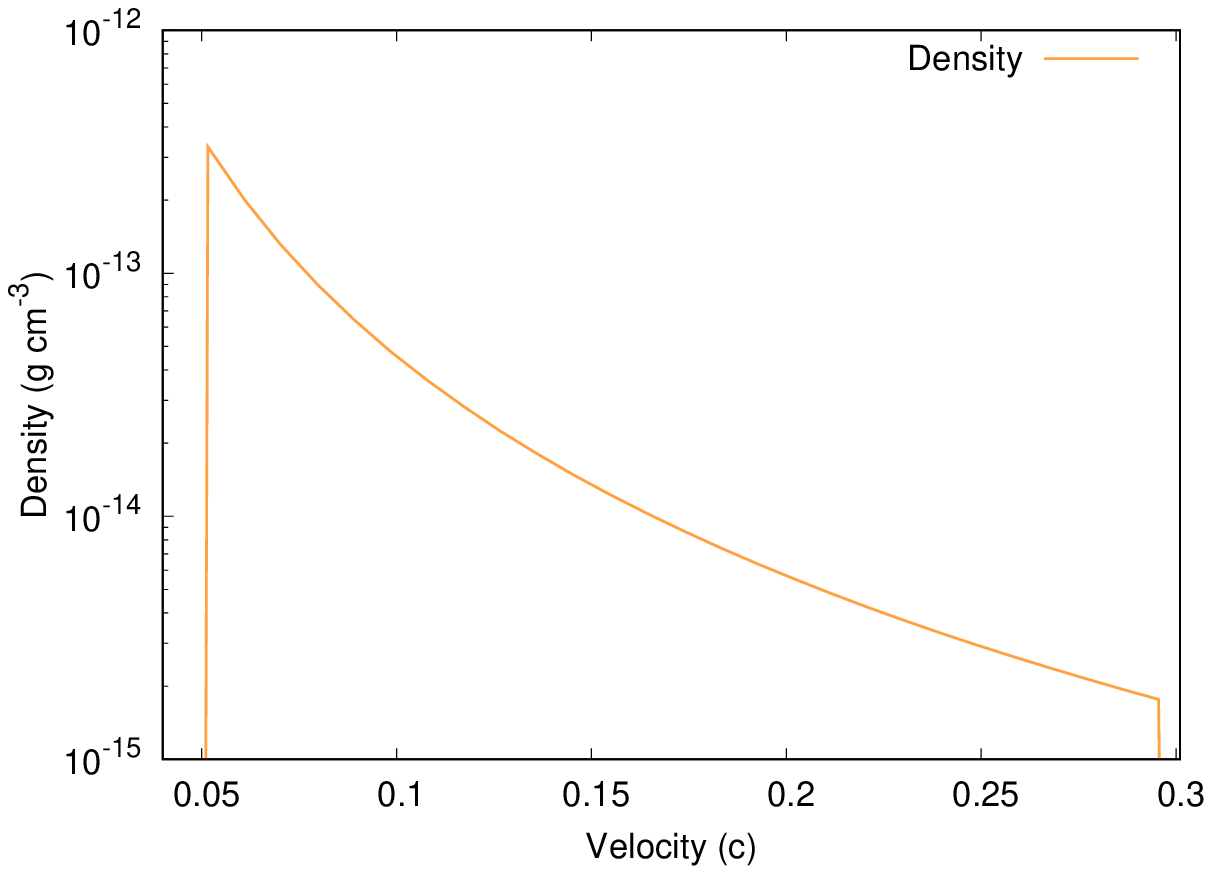} &
    \includegraphics[scale=0.7]{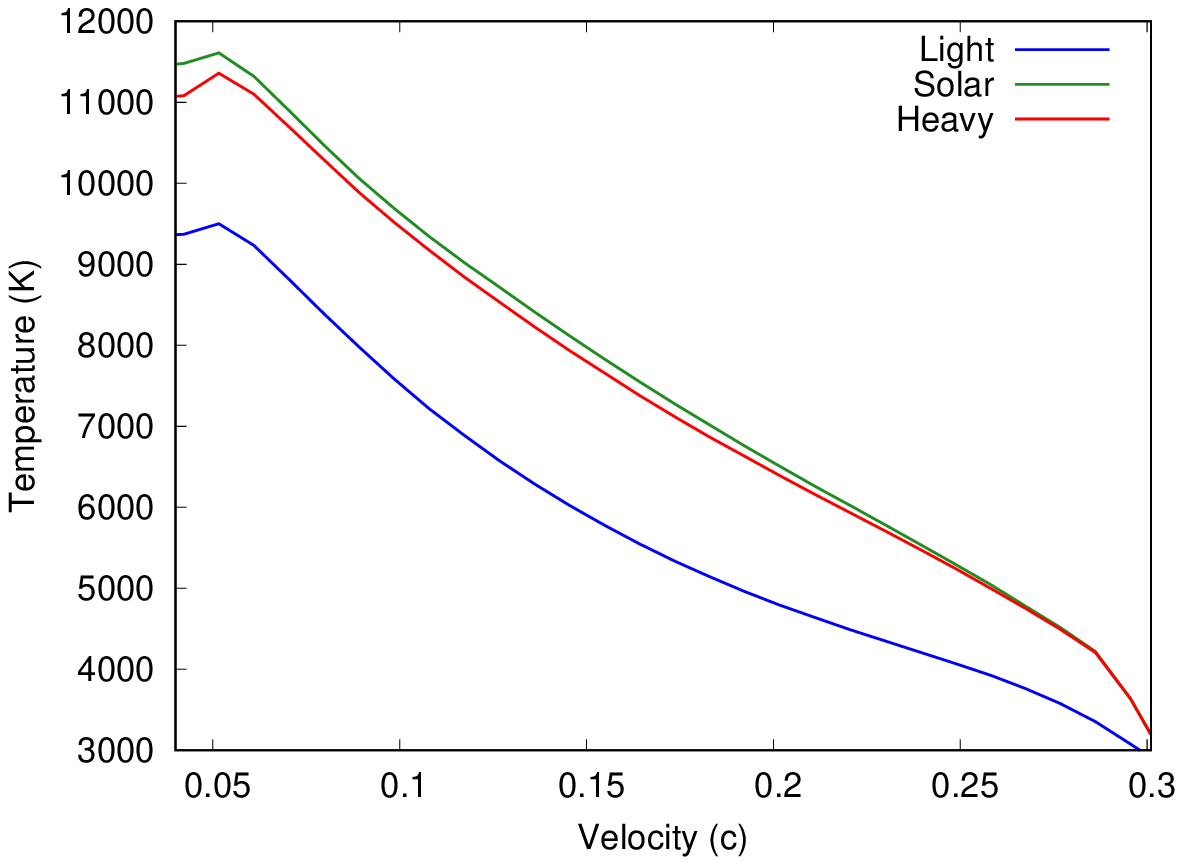}  \\
    \end{tabular}
\caption{
  \label{fig:structure}
  Density (left) and temperature (right) structures of the ejecta as a function of velocity at $t=1.5$ days after the merger. 
  The density structure is same for the three models. 
}
\end{center}
\end{figure*}

Figure \ref{fig:flux} shows our results of synthetic spectra at $t=$ 1.5 days after the merger. 
To show the contribution of different elements, we also plot the Sobolev optical depths in the ejecta at $v=0.2\ c$. 
The wavelengths of lines are blueshifted according to $v=0.2\ c$. 
This choice broadly captures the absorption features of the synthetic spectra, which indicates that the line forming regions in the ejecta are formed around this velocity for our models. 

The spectrum of AT2017gfo at $t=1.5$ days (gray curve) is also plotted, which is vertically shifted for comparison. 
Our synthetic spectra do not necessarily reproduce the overall spectral shape: 
the L model underproduces the NIR flux while the S and H models underproduce the optical flux. 
This is probably due to the assumption of the homogeneous abundance distribution in our models. 
It appears that a multi-component model is needed to reasonably reproduce the observed spectrum of AT2017gfo (\eg \citealp{Kawaguchi2018}). 
In this paper, therefore, we mainly focus on spectral features made by some elements. 

We find that Sr II and Ca II produce absorption lines at $\lambda\sim$ 8000 {\AA} and $\lambda\sim$ 6500 {\AA}, respectively, in the spectrum of the L model (arrows in the top panel of Figure \ref{fig:flux}). 
This is reasonable since these triplet lines are stronger than other lines as shown in Section \ref{sec:line result}. 
The Sr II absorption feature at $\lambda\sim$ 8000 {\AA} is consistent with that seen in AT2017gfo, which is blueshifted according to $v=0.2\ c$. 
This also supports the identification of this feature attributed to the Sr II lines \citep{Watson2019}.\par

For the S and H models, however, the Ca II and Sr II line features are overwhelmed by another broad absorption at around $\lambda\sim$ 7500 {\AA} (arrows in the middle and bottom panels of Figure \ref{fig:flux}), which is mainly caused by Ce III ($Z=58$), Tb III ($Z=65$) and Th III ($Z=90$). 
The highest contribution comes from the Tb III and Th III lines
(the rest frame wavelength is 9115.425 {\AA} for Tb III and 9910.075 {\AA} for Th III), 
while the Ce III lines give a minor contribution. 
This is mainly due to the large fraction of heavy elements in the S and H models.
Additional effects by the temperature are discussed in Section \ref{sec:Sr}.

\section{Discussion}
\label{sec:discussion}
\subsection{Identification of strontium and other heavy elements}
\label{sec:Sr}
\begin{figure}[th]
  \begin{center}
    \includegraphics[scale=0.7]{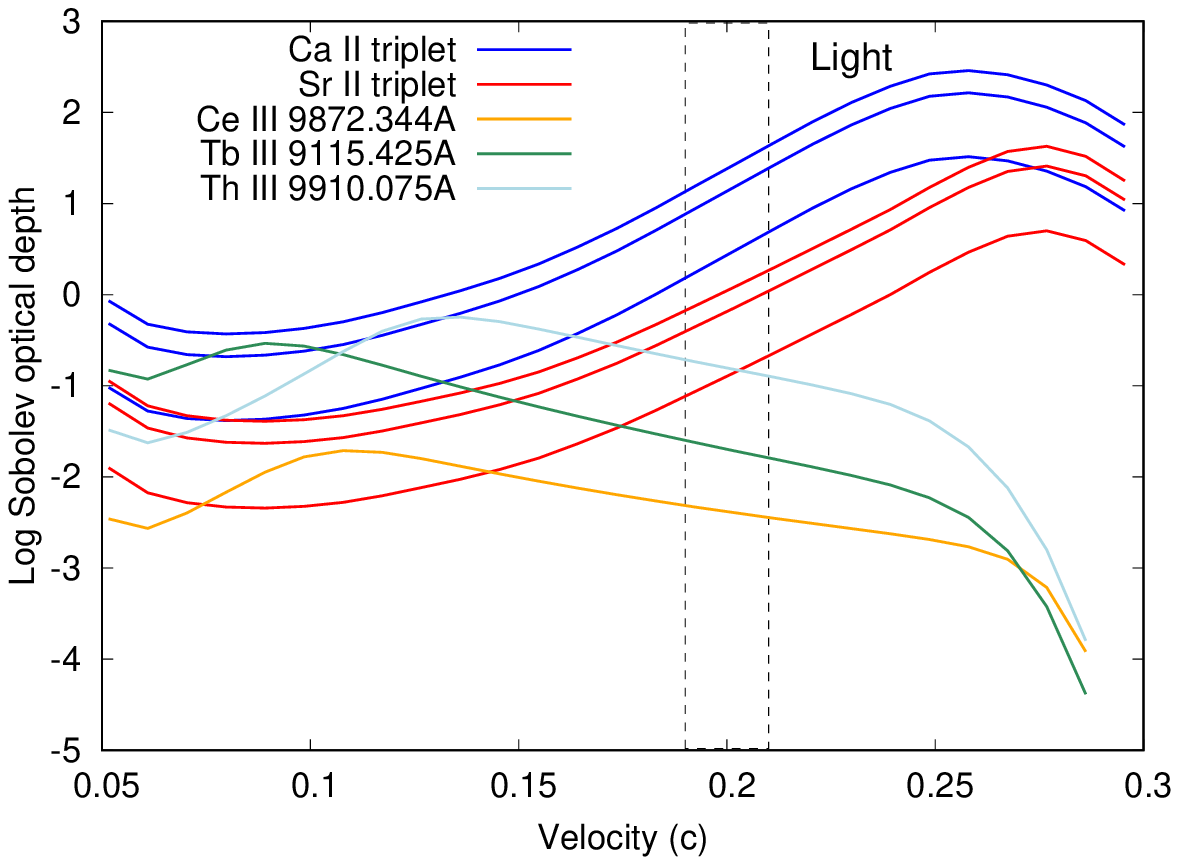} \\
    \includegraphics[scale=0.7]{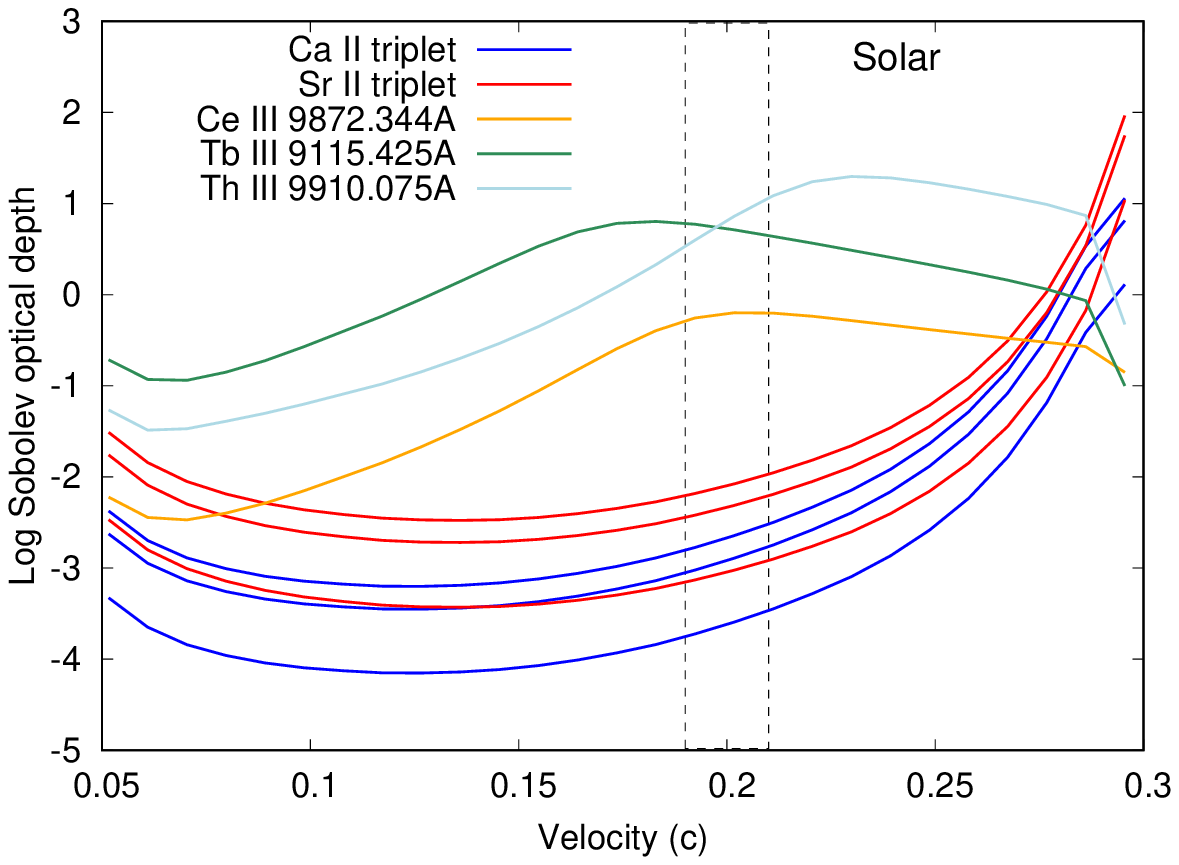}  \\
    \includegraphics[scale=0.7]{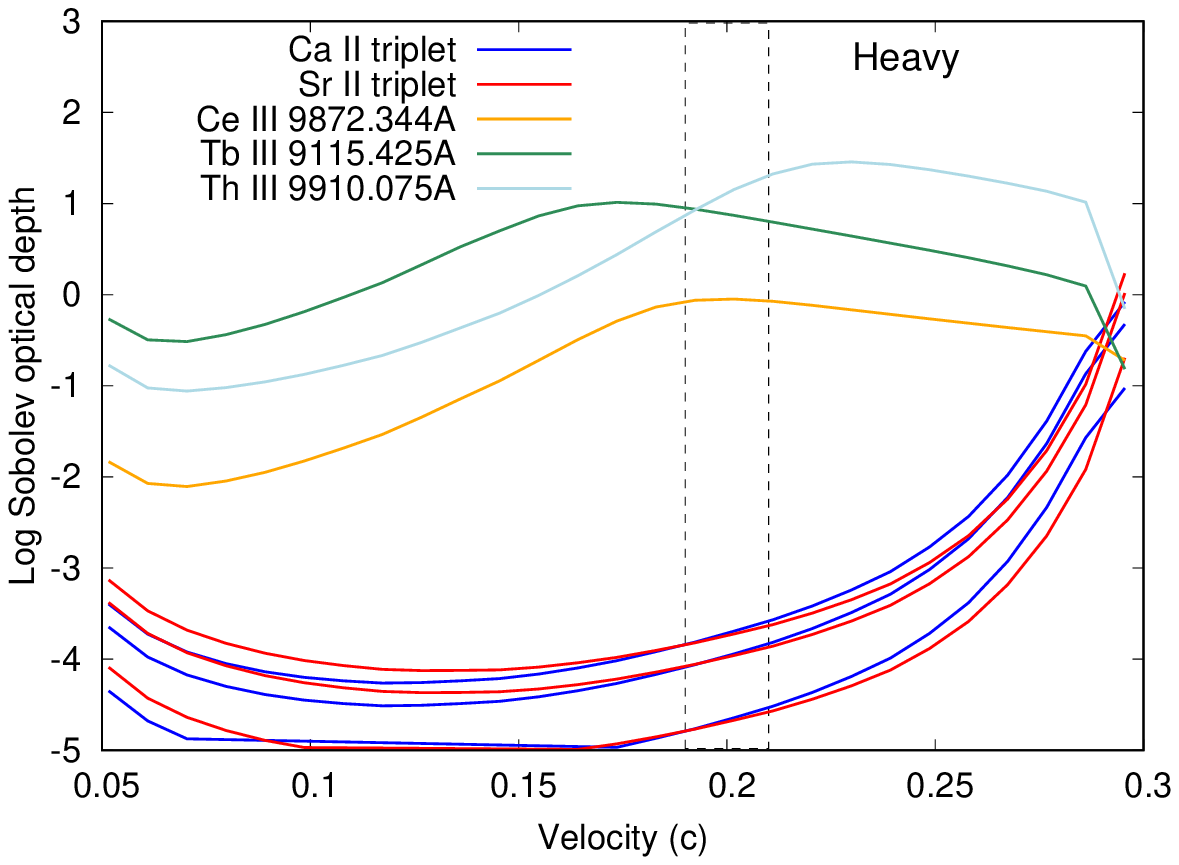}
\caption{
  \label{fig:v-tau}
  Distribution of the Sobolev optical depths as a function of velocity for the representative lines at $t=1.5$ days after the merger. 
  The top, middle and bottom panels show the results for the L, S and H models, respectively. 
  The rectangle region (dashed line) in each panel indicates $v\sim 0.2\ c$.
}
\end{center}
\end{figure}

Our results confirm that the absorption feature at $\lambda\sim$ 8000 {\AA} in the spectrum of AT2017gfo can be caused by the Sr II triplet, as reported by \citet{Watson2019}. 
However, it is emphasized that the Sr II lines are not always strongest: they give the strong absorption feature
only in our lanthanide-poor (L) model but not in our lanthanide-rich (S or H) models.
This suggests that the line forming region of AT2017gfo at $t=1.5$ days was lanthanide-poor (as discussed below).

Note that, in \citet{Watson2019}, the absorption due to the Sr II triplet is strongest at $\lambda \sim 8000$ {\AA} regardless of the existence of heavy elements. 
As the strength of the Sr II lines is sensitive to the temperature (see Figure \ref{fig:map}), this difference seems to be caused by the treatment of temperature in the ejecta.
Our simulations take the structure of density and temperature in the ejecta into account, and calculate the temperature from the radioactive heating rate in a consist way.
The temperature in the ejecta at $v=0.2\ c$ is $T\sim 5200\ {\rm K}$ for the L model, and it is $T > 7000\ {\rm K}$ for the S and H models. 
This difference is mainly caused by the higher heating rate for the latter models. 
Since these models synthesize heavier elements than those in the L model, the average radioactive heating rate is higher \citep{Wanajo2018a}.
Also, when the ejecta have more heavy elements, the overall opacity becomes higher, which suppress the radiative cooling (\citealp{Kasen2013}; \citealp{TanakaHotokezaka2013}; \citealp{Tanaka2018}). 
Therefore, in the S or H models, the temperature in the ejecta is higher than that in the L model for a fixed velocity.

To see the effect of abundance distribution and temperature to the line strength, we show the density and temperature structures in the ejecta in Figure \ref{fig:structure} and the Sobolev optical depths for each model as a function of velocity in Figure \ref{fig:v-tau}. 
When we focus on the Sr II triplet lines, it is clear that they are strong in the L model. 
On the other hand, these lines are not necessarily strongest in the S and H models at the same velocity. 
Since the mass fraction of Sr is almost the same between the L and S models (see Table \ref{tab:abun}), the strength of Sr II lines is mainly affected by the temperature difference (see also Figure \ref{fig:map}).
Also, provided that the temperature is lower ($T\sim 5000$ K) in the S model, the Sr II line strength becomes comparable to those of heavy elements, as shown in Figure \ref{fig:tau}.
Thus, the feature by the Sr II lines does not appear in the spectra of lanthanide-rich ejecta.

For the line strength of heavy elements, the abundance distribution plays a major role.
As shown in Figure \ref{fig:tau}, the strength of Ce III, Tb III, and Th III lines is not largely changed between $T=5000$ K and $T=7000$ K and they are comparable or even stronger than other lines.
Therefore, if the ejecta have a large amount of heavy elements as in the S model, the spectra show the feature of these heavy elements.

The optical depths in the S and H models are qualitatively similar, and thus, the synthetic spectra of these models also become similar. 
The strength of Ca II and Sr II lines for the H model is weaker than that for the S model because of their smaller abundances (bottom panel of Figure \ref{fig:v-tau}).
Note that a significant difference between the spectra for the S and H models can appear at the later phase when the density and temperature in the ejecta decrease (see also Section \ref{sec:NIR}). 

Our results suggest that the absorption features depend on temperature as well as abundance distribution in the ejecta. 
This means that different spectral features can appear in future kilonovae depending on, \eg the ejecta masses and abundance distribution.
In fact, our S and H models show spectra with absorption features by heavy elements such as Ce, Tb and Th. 
The lines of these doubly ionized elements have been identified in chemically peculiar stars by using theoretical $gf$ values (\eg Ce III: \citealp{Ryabchikova2006}, Th III: \citealp{Ryabchikova2007}). 
Identification of such elements provides the direct evidence that heavy $r$-process elements are indeed synthesized in NS merger ejecta. 

It should be noted that the transition probabilities of the doubly ionized ions of heavy elements are rather uncertain. 
The transition probabilities of these lines have been obtained only by theoretical calculations (Ce III: \citealp{Biemont2002b}, Th III: \citealp{Biemont2002a}, Tb III: Database on Rare Earths At Mons University, DREAM; \citealp{Biemont1999}), 
while those for singly ionized Ca and Sr are experimentally evaluated (\eg Ca II: \citealp{Ca}; Sr II: \citealp{Sr}). 
Since kilonova ejecta are expected to have a wide range of temperature and ionization states, more experimental and observational calibrations for transition probabilities are necessary (see also Section \ref{sec:NIR}). 

It is also cautioned that our models are calculated by assuming  one-dimensional ejecta with homogeneous abundance distribution.
It is not clear about how multi-dimensional modeling will affect our results presented here, depending on the spatial distributions of, \eg ejecta masses, velocities, and abundances. 
We leave such exploration to future work.

\subsection{Strontium and calcium as a tracer of high $\Ye$ ejecta}
\label{sec:Ca}
\begin{figure}[htp]
  \begin{center}
    \includegraphics[scale=0.7]{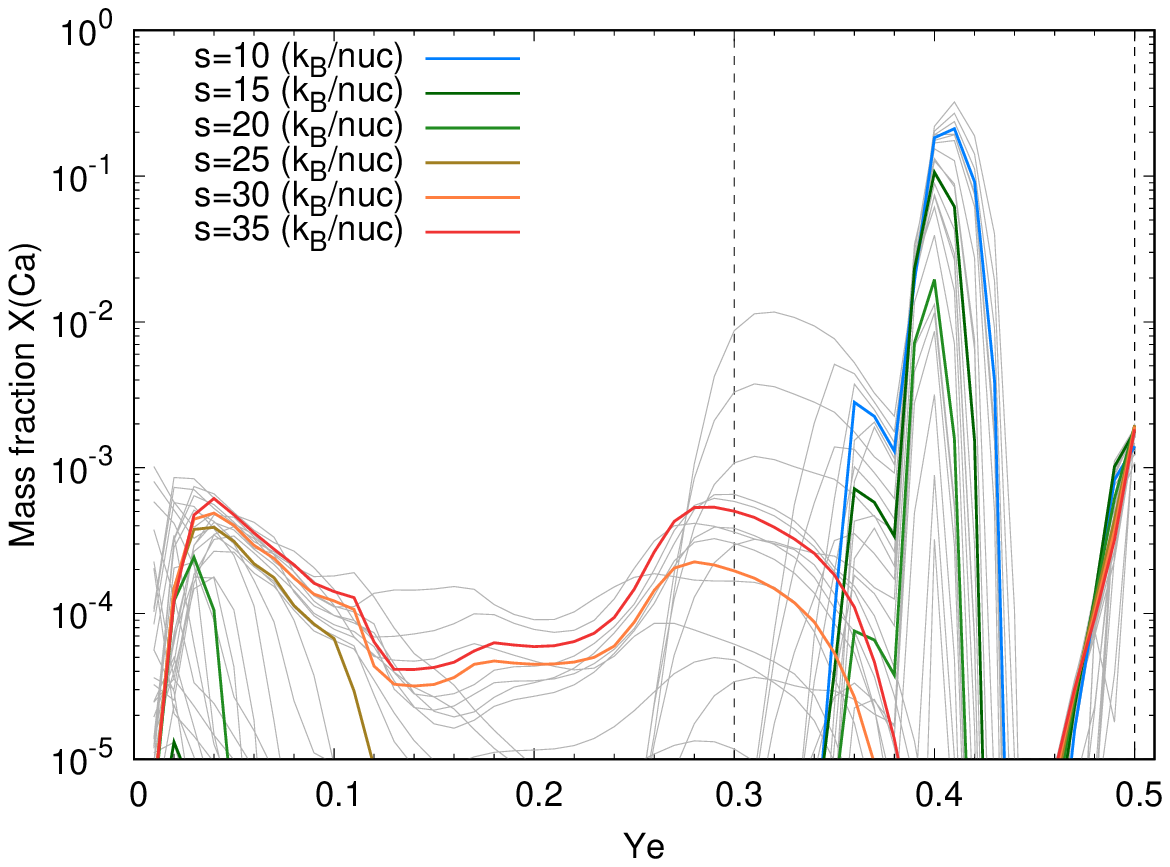}  \\
    \includegraphics[scale=0.7]{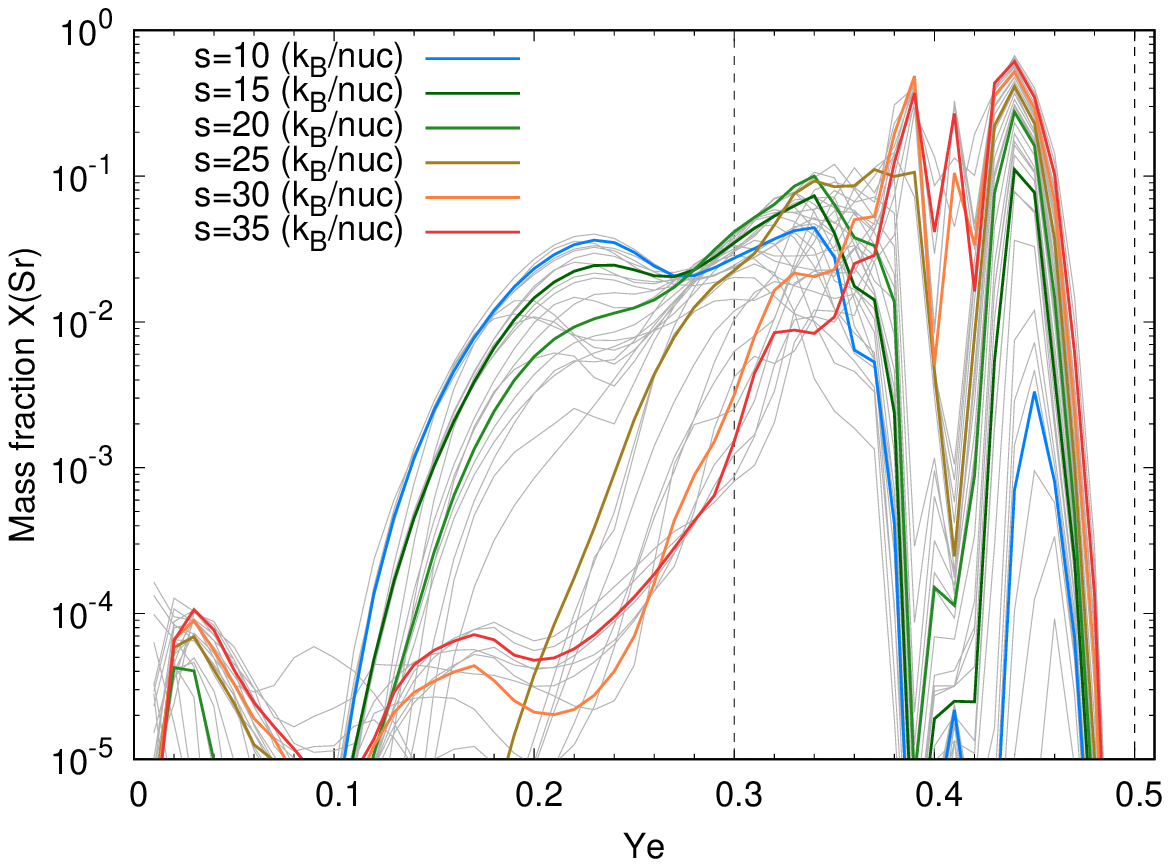}  \\
    \includegraphics[scale=0.7]{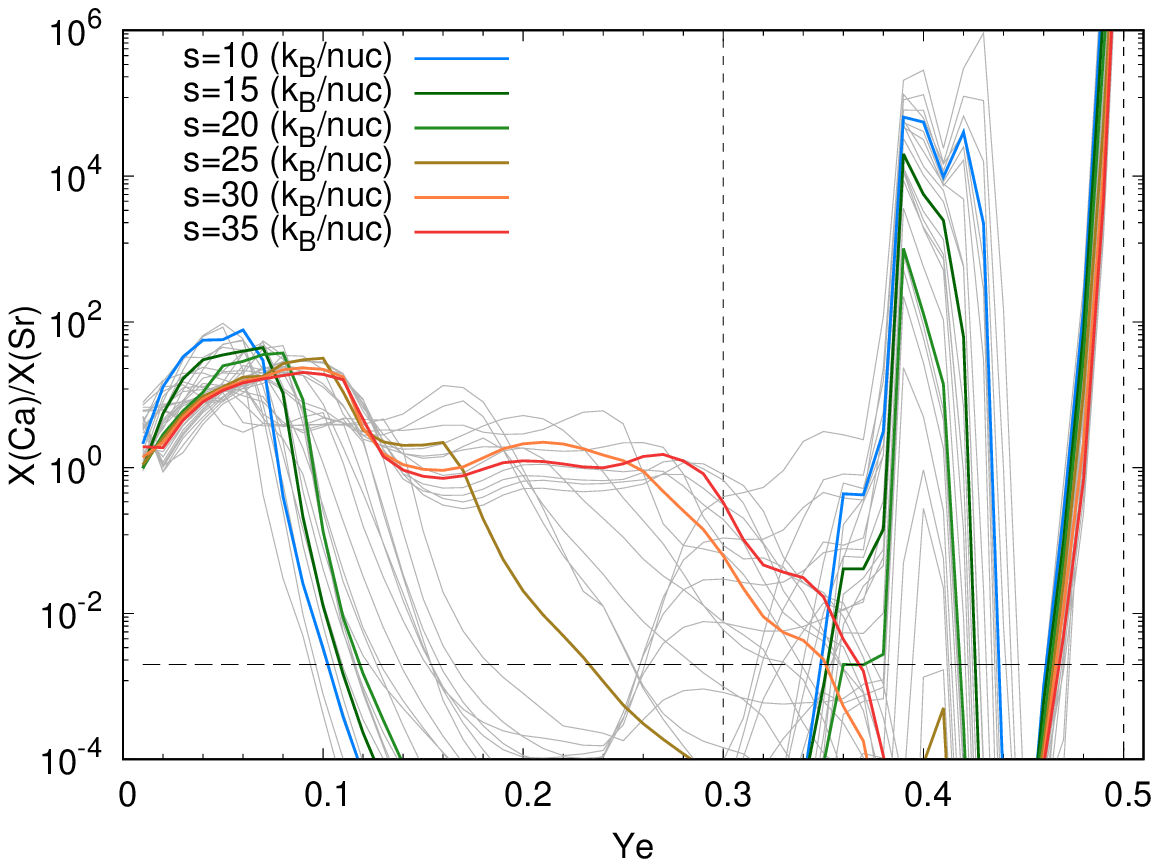}  \\
\caption{
  \label{fig:mass-ratio}
  Mass fractions of Ca (top) and Sr (middle) as a function of $\Ye$
  for the individual outflows with the various sets of velocity and entropy (gray lines). 
  Colored lines show the results with $v=0.2\ c$ and the six different entropies in the legend. 
  The bottom panel shows the ratio of the mass fraction of Ca to that of Sr.  
  The horizontal dashed line indicates X(Ca)/X(Sr)$=0.002$, below which the absence of the Ca line in AT2017gfo can be explained.
  The Ca abundance is dominated by $^{48}$Ca except for $\Ye\gtrsim 0.45$.
}
\end{center}
\end{figure}
\begin{figure}[ht]
  \begin{center}
    \includegraphics[scale=0.7]{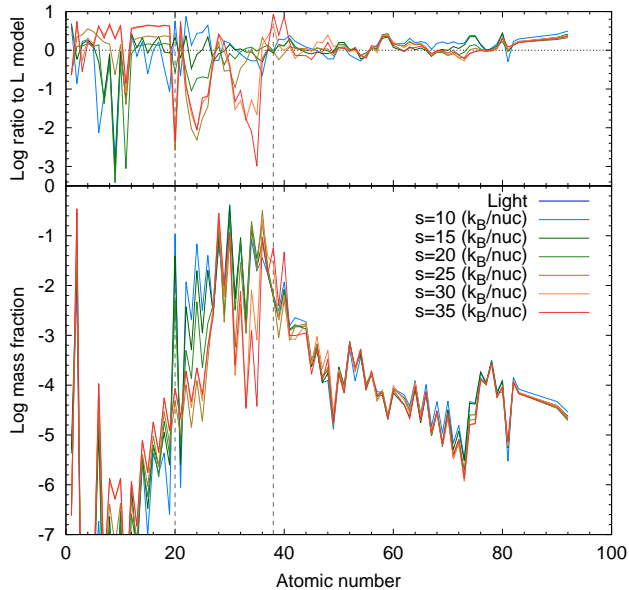}
\caption{
  \label{fig:abun-entropy}
  Final abundances for the individual outflows with the fixed velocity of $v=0.2\ c$ and the six different entropies in the legend as a function of atomic number, compared to those for the L model. 
  The top panel shows the ratio of abundances for each outflow to those for the L model. 
  Dashed lines indicate Ca ($Z=20$) and Sr ($Z=38$). 
  Abundances are not largely changed among these cases except for the range around $Z=20$--40. 
}
\end{center}
\end{figure}

Our results suggest that the Sr II and Ca II triplet lines can be used as a tracer of high-$\Ye$ ejecta.
The L model shows a strong absorption feature of the Ca II lines, while the spectra of AT2017gfo does not exhibit such features. 
This implies that the abundance of Ca in the line forming region of AT2017gfo is smaller than that in our L model. 
We perform the same radiative transfer simulations for the L model by varying the mass fraction of Ca to reconcile our model with the case for AT2017gfo. 
The result indicates that the Ca II lines disappear when the ratio X(Ca)/X(Sr) is smaller than $\sim 0.002$. \par

To infer the relevant physical properties from this constraint, we show the mass fractions of Ca and Sr as a function of $\Ye$ for the individual outflows (Section \ref{sec:line method}) with different velocities and entropies in Figure \ref{fig:mass-ratio}.
We first focus on the range of $\Ye = 0.30$--0.50 as a relevant condition in the L model (Figure \ref{fig:hist ye}) and the velocity of only $v=0.2\ c$ which is consistent with the blueshift of the Sr II triplet in AT2017gfo. 
For these conditions, the Sr abundance (dominated by $^{88}$Sr) is always relatively high (color lines in the middle panel), which is produced both in nuclear equilibrium and by neutron capture over a wide range of $\Ye$. 
By contrast, Ca, which is dominated by $^{48}$Ca for $\Ye < 0.45$, is preferentially produced in nuclear equilibrium with $\Ye$ near its proton-to-nucleon ratio of 20/48 $=$ 0.417 \citep{Meyer1996a, Wanajo2013}. 
For this reason, the condition of X(Ca)/X(Sr) $<$ 0.002 can be achieved only with the relatively high entropies of $s \ge 25 k_{\rm B}/{\rm nuc}$ (bottom panel), which inhibit nuclear equilibrium to be established. 
This implies that the blue component of AT2017gfo, which produces the Sr absorption line, comes from relatively high entropy ejecta. 
Such a condition may be realized in the viscously heated, post-merger ejecta (\eg \citealp{Just2015, Lippuner2017, Fujibayashi2020}) 
or in the shocked dynamical ejecta along the polar direction (\eg \citealp{Wanajo2014, Sekiguchi2015, Radice2018}). \par

It should be noted that the conditions discussed above ($v \sim 0.2\ c$ and s $\ge 25 k_{\rm B}$/nuc) differ from the velocity and entropy distributions for our L model (the top panels of Figure \ref{fig:hist appendix} in Appendix \ref{sec:appendix2}). 
However, this does not substantially affect our results shown in Section \ref{sec:spectra} and \ref{sec:Sr} except for the Ca features. 
As cautioned in \citet{Wanajo2018a}, our model obtained by fitting a given reference abundance distribution is not necessarily the unique solution. 
We show the final abundances for the individual outflows with the fixed velocity of $\sim 0.2\ c$ and the entropy of $\sim$ 10--25 $k_{\rm B}$/nuc in Figure \ref{fig:abun-entropy}. 
It is clear that a similar fit to that for the L model can be obtained with those parameters, except for the range around $Z=20$--40. 
The distributions of $\Ye$ for those entropies are similar to that for the L model, which tend to, however, slightly shifted to the higher-$\Ye$ side for a higher entropy. 

Note also that we assume a solar $r$-process-like abundance distribution over a given range of atomic mass numbers to obtain our models (Section \ref{sec:line method}). 
If we relaxed this constraint, the ratio of X(Ca)/X(Sr) $<$ 0.002 would also be obtained with s $\sim 10 k_{\rm B}$/nuc and $\Ye \lesssim 0.35$ (Figure \ref{fig:mass-ratio}, bottom panel). 

Our results imply that the Ca triplet line can be observed in future NS merger events. 
In particular,the post-merger ejecta with relatively high $\Ye$ and relatively low entropy may have high $^{48}$Ca abundance \citep{Fujibayashi2020}. 
For such a case, we expect the Ca absorption feature with a moderate blueshift since the post-merger ejecta have relatively low expansion velocity ($v \sim 0.05\ c$). 
The detection of the Ca line in a future event will provide unique information on the ejecta condition as examined above for AT2017gfo.

Finally, it is interesting to note that $^{48}$Ca is one of the isotopes whose origin remains a mystery. 
This isotope is synthesized most efficiently in the conditions of $\Ye \sim 0.40$--0.42 and $\lesssim 15 k_{\rm B}$/nuc (\citealp{Meyer1996a}; \citealp{Wanajo2013}). 
Such a condition may be achieved in high-density type Ia supernovae \citep{Woosley1997a}, electron-capture supernovae \citep{Wanajo2013, Jones2019a}, low-mass core-collapse supernovae \citep{Wanajo2018b}, collapsars \citep{Fujibayashi2020b}, or NS mergers \citep{Wanajo2018a, Fujibayashi2020}. 
If the absorption line caused by the Ca II triplet appears in future kilonova spectra, we will be able to confirm the production of $^{48}$Ca by NS mergers. \par

\subsection{Importance of the NIR line list}
\label{sec:NIR}
\begin{figure}[t]
    \includegraphics[scale=0.66]{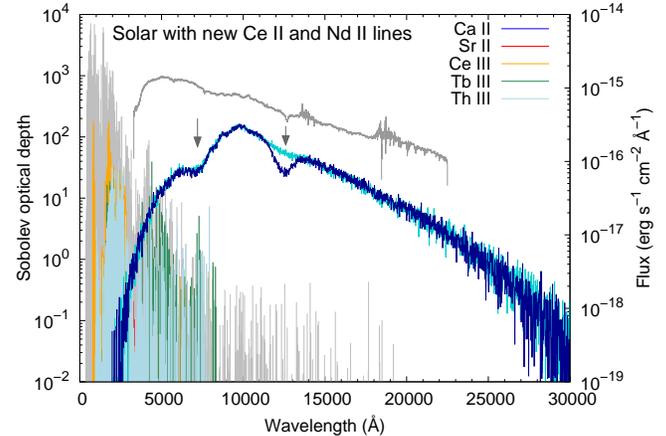}
\caption{
  \label{fig:flux Ce}
  Same as the middle panel of Figure \ref{fig:flux} (the S model) but with the additional transitions for Ce II \citep{Cunha2017} and Nd II \citep{Hasselquist2016} in the NIR wavelengths. 
  The spectrum in light-blue color is that without the additional lines. 
}
\end{figure}

The spectra of AT2017gfo show absorption features in the NIR region (\eg \citealp{Pian2017}). 
We have performed systematic calculations and radiative transfer simulations including the NIR lines, but it is difficult to unambiguously identify the absorption features. 
This is because the transition data are not sufficient in the NIR wavelengths compared with those in the UV to optical wavelengths. 
Atomic data for the NIR transitions have been limited mainly due to a lack of high-resolution stellar spectroscopic observations and laboratory experiments in the NIR wavelengths. 

Recently, new measurements become available owing to the development of NIR spectroscopy such as the APOGEE survey \citep{APOGEE2008}. 
When the abundances in certain Galactic stars have been obtained from their optical spectra, one can estimate the transition probabilities for NIR lines by measuring these stars in the NIR wavelengths. 
In fact, two such analyses about Ce II ($Z=58$) and Nd II ($Z=60$) have been performed for the APOGEE survey, which give the transition probabilities for the 9 lines of Ce II \citep{Cunha2017} and 10 lines of Nd II \citep{Hasselquist2016} in $H$-band. 

We perform here the same simulations as in Section \ref{sec:spectra} but by including these new lines. 
The synthetic spectrum for the S model with the new Ce II and Nd II lines is shown in Figure \ref{fig:flux Ce}. 
The result shows an additional absorption feature at $\lambda \sim$ 13000 {\AA}, which is caused by the new Ce II lines. 
These Ce II lines are strong at around $T\sim4000$--5000 K similar to the Sr II lines. 
The H model also shows the same absorption feature, while the L model does not show it because of a small fraction of lanthanides (Table \ref{tab:abun}). 
Due to incompleteness of the line list in the NIR wavelengths, there is always a possibility that other unknown lines show stronger absorption lines.  
Therefore, it is not evident if the Ce II lines can be observed in future (S or H model-like) events. 
Our analysis, however, demonstrates that atomic data in the NIR wavelengths are important to fully decode kilonova spectra in particular at the later phase when kilonovae are brighter in NIR.

\section{Conclusions}
\label{sec:conclusion}
We have performed the systematic calculations of line strength for bound-bound transitions and radiative transfer simulations toward element identification in kilonova spectra. 
We have found that Sr II triplet lines appear in the spectrum at $v=0.2\ c$ for our lanthanide-poor (L) model, which is consistent with the feature at $\lambda \sim$ 8000 {\AA} in the spectrum of GW170817/AT2017gfo. 
We also have found that Ca II triplet lines can appear in kilonova spectra. 
This is due to the fact that these two elements are co-produced in relatively high $\Ye$ ejecta, and their ions have similar atomic structures: 
only one electron in the outermost shell and the high transition probabilities for bound-bound transitions. 
We have shown that line strength strongly depends on the abundance distribution and temperature in the ejecta. 
For our lanthanide-rich (S or H) model, the spectra show the features of doubly ionized heavier elements, such as Ce, Tb and Th. 
This points to a possibility that we can obtain the evidence of production of such heavy $r$-process elements in future NS merger events. 

Since our results for the lanthanide-poor (L) model account for the presence of Sr II lines in the spectrum of GW170817/AT2017gfo, the line forming region in the ejecta of GW170817 was likely dominated by relatively high $\Ye$ ($> 0.30$) material. 
Furthermore, we can constrain the physical condition of the ejecta by using the strength of Ca II and Sr II lines. 
Absence of the Ca II absorption in the spectra of GW170817/AT2017gfo implies that the X(Ca)/X(Sr) ratio should be less than 0.002. 
This ratio can be achieved in the NS merger ejecta with relatively high entropy ($s \gtrsim 25\ k_{\rm B}$/nuc) for the velocity of $\sim 0.2\ c$, which may be realized in the post-merger ejecta or in the shock heated dynamical ejecta. 
Since a high mass fraction of Ca (dominated by $^{48}$Ca) is expected only with $\Ye \gtrsim 0.40$, the Ca II triplet can be used as a tracer of high-$\Ye$ ejecta.  
We speculate that mildly Doppler-shifted Ca II lines from the high-$\Ye$ post-merger ejecta with relatively small velocity will be identified in future events. 

It is still challenging to identify the NIR absorption features in kilonova spectra because the atomic data are still incomplete in the NIR wavelengths. 
We have shown that the newly measured transitions of Ce II lines from the APOGEE survey affect the synthetic spectra in the NIR wavelengths.
Although we cannot conclude that these lines will be certainly visible in future events, our study demonstrates the importance of updating the NIR line list to fully decode the spectra of kilonovae.

\acknowledgements
We thank S. Fujibayashi for fruitful discussion on the physical conditions of NS mergers. 
This research was supported by the Grant-in-Aid for Scientific Research from JSPS (19H00694, 20H00158, 18H05859) and MEXT (17H06363).

\appendix

\section{Effects of the line list to the light curve}
\label{sec:appendix1}
\begin{figure}[bht]
  \begin{center}
  \begin{tabular}{cc}
    \includegraphics[scale=0.7]{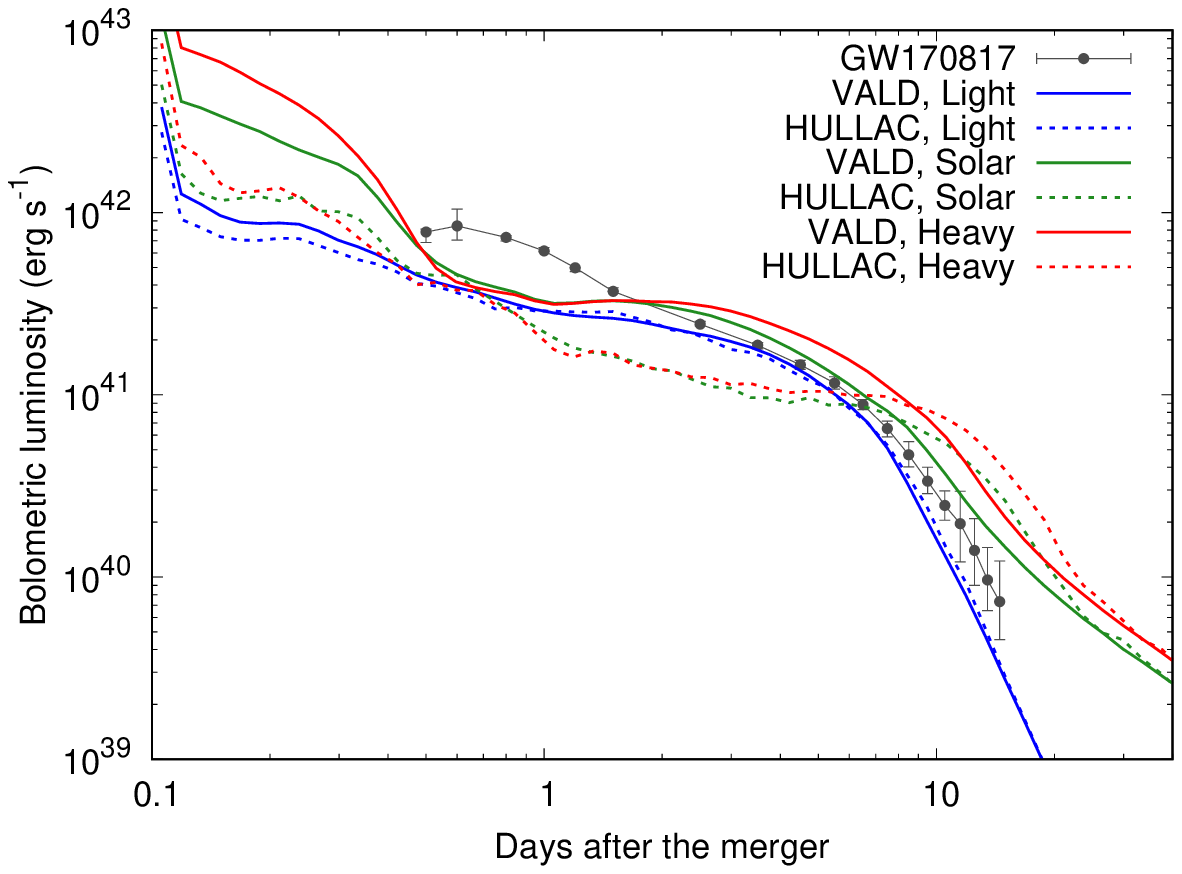}
    \includegraphics[scale=0.7]{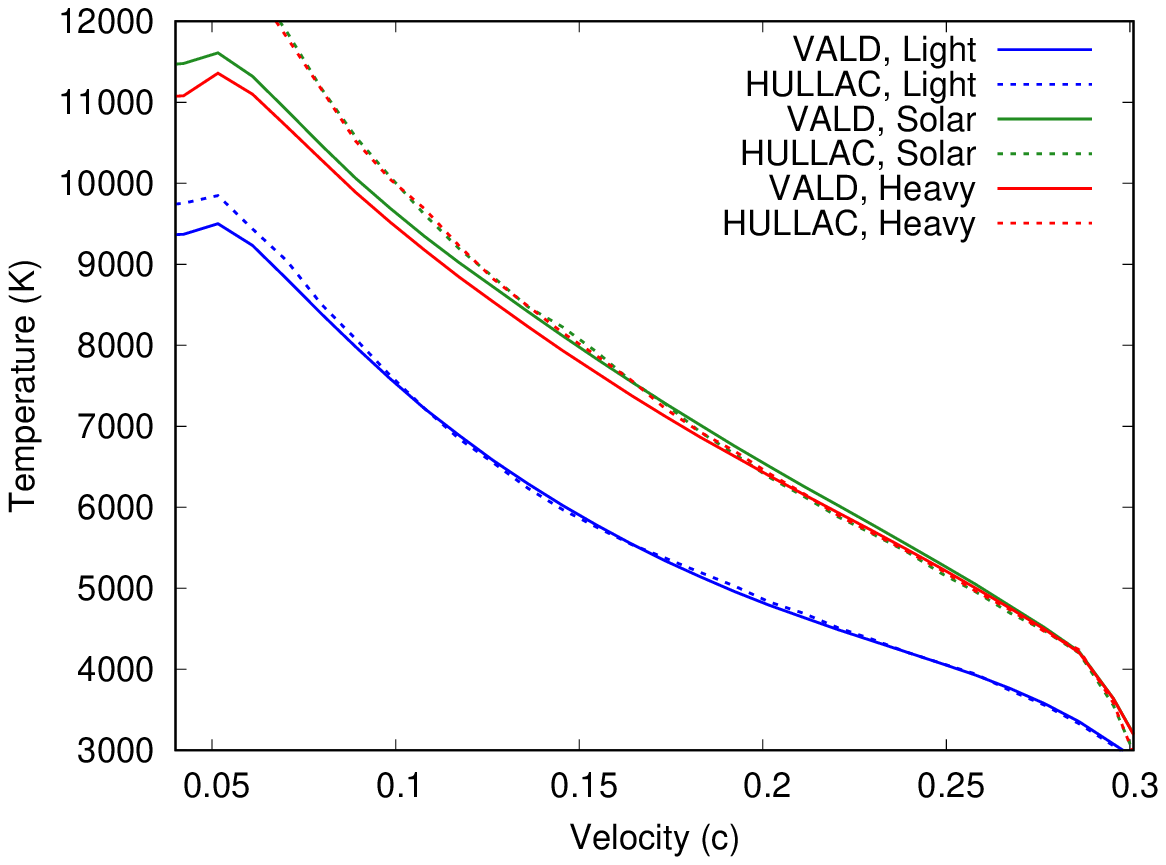}
  \end{tabular}
\caption{
  \label{fig:Lbol}
  Left: Bolometric light curves calculated with the two line lists. 
  The bolometric light curve of GW170817 (the dots with error bars connected by lines, \citealp{Waxman2018}) is also shown for comparison. 
  Right: Radial distribution of temperature at $t=$ 1.5 days. 
  In both panels, the solid and dashed lines show the results with the line lists from the VALD database \citep{Piskunov1995,Kupka1999,Ryabchikova2015} and HULLAC calculations \citep{Tanaka2020}, respectively.
  Different colors represent different models as shown in the legend.
  }
\end{center}
\end{figure}

We inspect the impact of the adopted line list to the ejecta properties. 
The line list from the VALD database is not necessarily complete, in particular, for heavy elements in the NIR wavelengths.
Thus, this incompleteness may affect the physical condition of the ejecta such as temperature. 
Here, we perform radiative transfer simulations as described in Section \ref{sec:spectra} with the theoretical line list from \citet{Tanaka2020}.
This line list was constructed by systematic atomic calculations using the HULLAC code. 
Transition data of this line list are not necessarily accurate in the relevant wavelengths, and thus, they are not suitable for the identification of lines in kilonovae. 
However, since the data coverage is complete, it may give a more reliable estimate of the total opacity in the ejecta. \par

We confirm that the incompleteness of the VALD data does not have a significant impact to the temperature structure of the ejecta.
The left panel of Figure \ref{fig:Lbol} shows the results of light curves using both line lists. 
For the L model, which has a small fraction of lanthanides, the light curves are almost the same for both line lists.
The light curves of the S and H models, which include a larger fraction of lanthanides, are more affected by the incompleteness of the line list: when the VALD database is used, our model underestimates the opacity, and thus, the luminosity tends to be higher. 
Nevertheless, the temperature structure of the ejecta at $t=1.5$ days after the merger is almost similar for all the models (the right panel of Figure \ref{fig:Lbol}). 
This is because the temperature in the ejecta is mainly controlled by the heating rate, while the luminosity largely depends on the opacity (\ie the line list). 
Because of the similarity of the temperature structure, we conclude that the incompleteness of the line list does not substantially affect our synthetic spectra at $t = 1.5$ days.

\section{Properties of models}
\label{sec:appendix2}

\begin{figure}[htb]
  \begin{center}
    \begin{tabular}{cc}
    \includegraphics[scale=0.65]{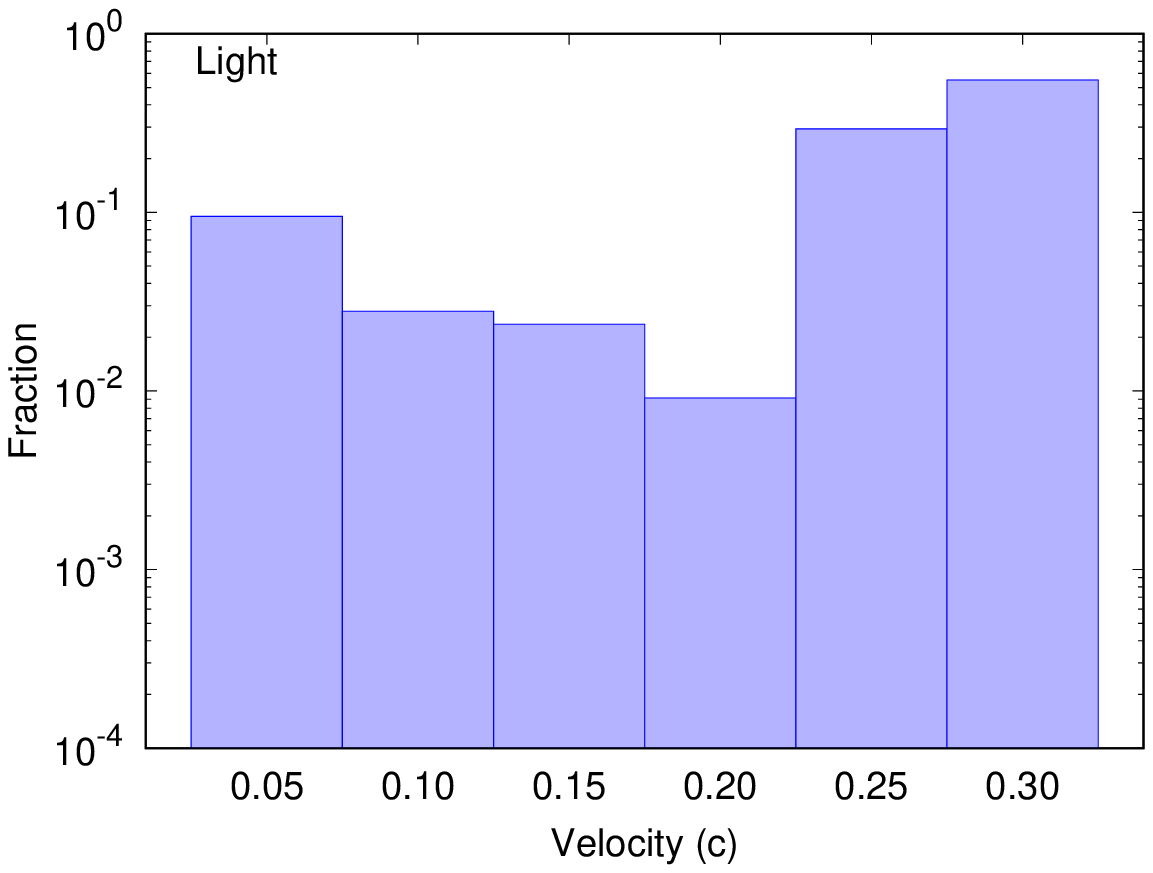} &
    \includegraphics[scale=0.65]{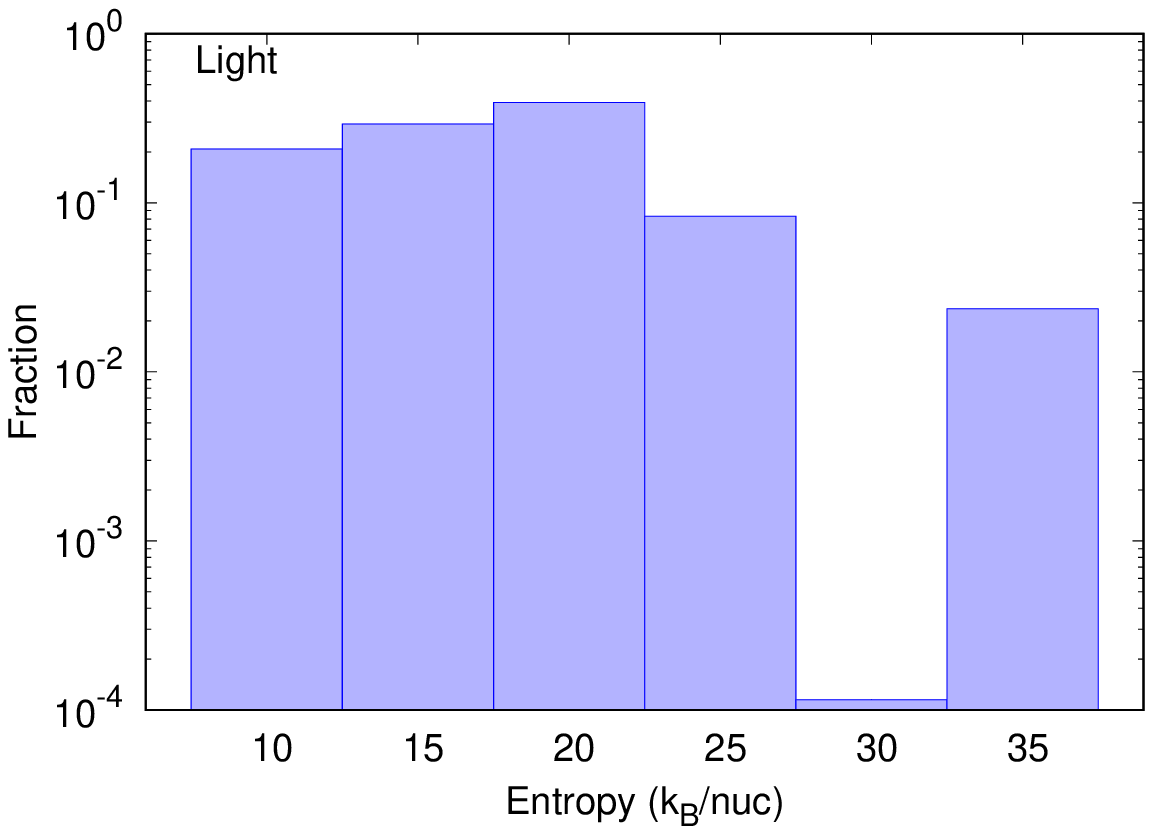} \\
    \includegraphics[scale=0.65]{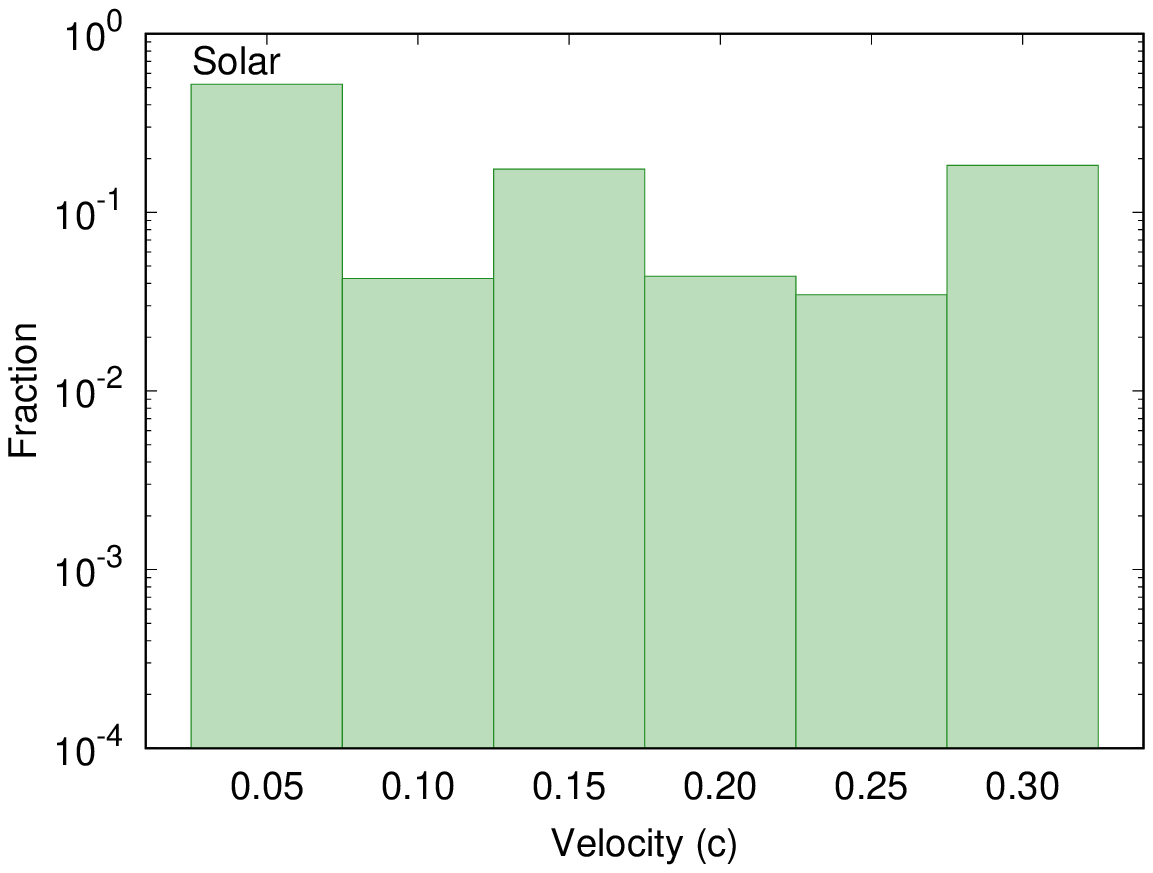} &
    \includegraphics[scale=0.65]{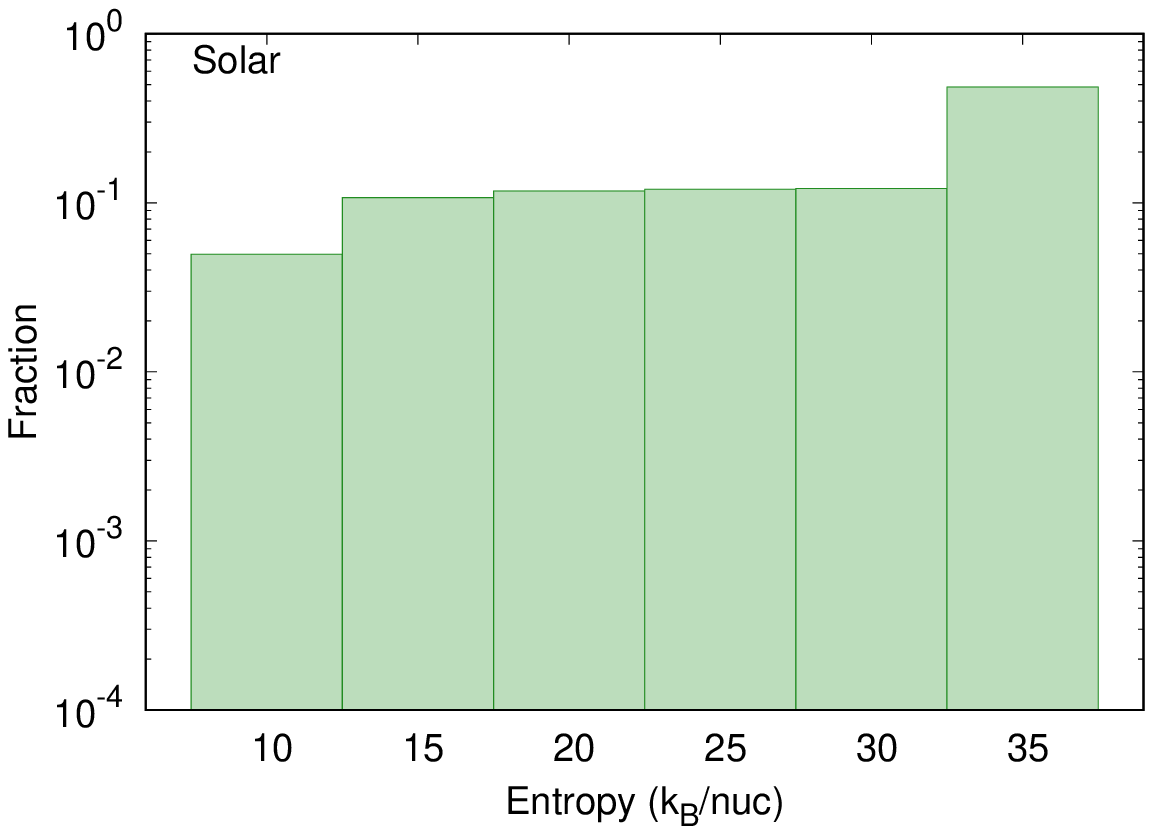} \\
    \includegraphics[scale=0.65]{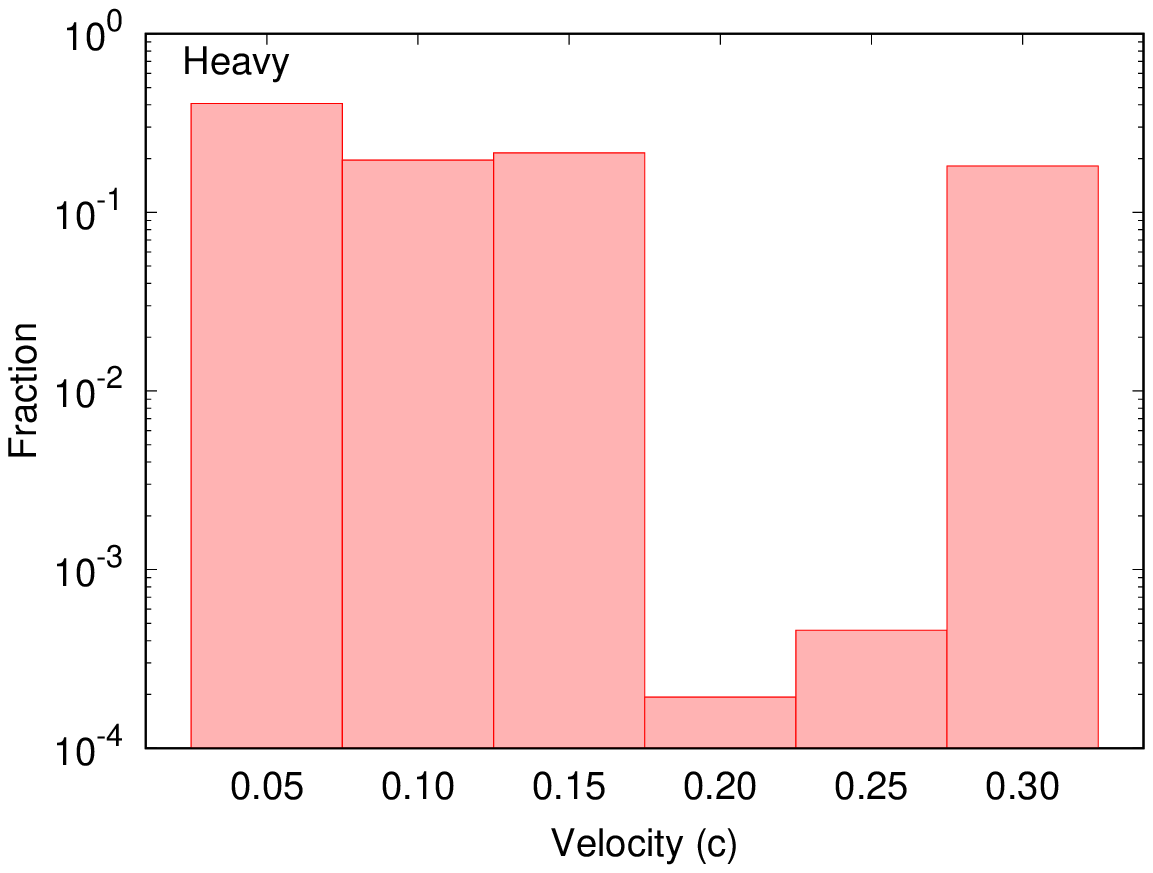} &
    \includegraphics[scale=0.65]{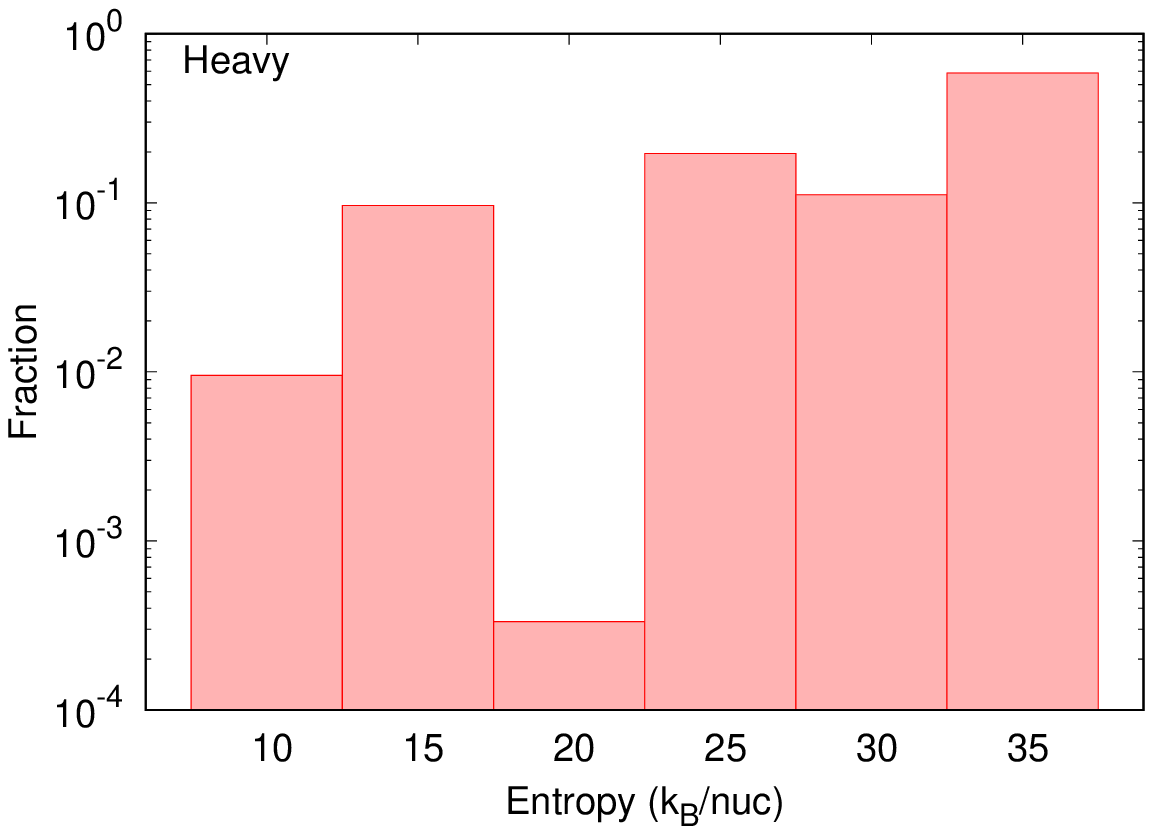}  \\
    \end{tabular}
\caption{
  \label{fig:hist appendix}
  Histograms of velocity (left) and entropy (right)
  for L (top), S (middle) and H (bottom) models. 
}
\end{center}
\end{figure}

We present the distributions of velocity and entropy for our L, S, and H models in Figure \ref{fig:hist appendix}. 
These distributions are not necessarily the unique solutions for these models as noted in Section \ref{sec:Ca}.

\end{document}